\RequirePackage[l2tabu, orthodox]{nag}
\RequirePackage{snapshot}

\documentclass{article}

\usepackage{amsthm,amsmath,amsfonts,amssymb}
\usepackage[english]{babel} %
\usepackage{subfigure}
\usepackage{setspace}
\usepackage{tabularx} %
\usepackage{ctable} %
\usepackage{enumerate}
\usepackage{epsfig}
\usepackage{graphics}
\usepackage{graphicx}
\usepackage{color}
\usepackage{booktabs}
\usepackage{longtable}
\usepackage{url}
\usepackage{placeins}

\usepackage[short,12hr]{datetime}
\usepackage{fullpage}
\usepackage{srcltx}
\usepackage{flushend}
\usepackage{mathptmx}

\usepackage[normalem]{ulem}

\usepackage{bm}

\def\E{\mathrm{I\! E}}

\allowdisplaybreaks[4] %
\newcommand{\R}{\mathbb{R}}

\newcommand{\barma}{${\beta}$ARMA}

\DeclareMathAlphabet\mathbfcal{OMS}{cmsy}{b}{n}

\title{
Beta seasonal autoregressive moving average models}

\author{
F\'abio M. Bayer\thanks{F.~M.~Bayer
is with the
Departamento de Estat\'istica and LACESM,
Universidade Federal de Santa Maria, RS, Brazil,
e-mail: bayer@ufsm.br}
\and
Renato J. Cintra\thanks{R.~J.~Cintra
is with the
Signal Processing Group,
Departamento de Estat\'istica,
Universidade Federal de Pernambuco, PE, Brazil,
e-mail: rjdsc@de.ufpe.br
}
\and
Francisco Cribari-Neto\thanks{F.~Cribari-Neto
is with the
Departamento de Estat\'istica,
Universidade Federal de Pernambuco, PE, Brazil,
e-mail: cribari@de.ufpe.br
}
}
\date{}

\begin{document}

\maketitle

\doublespacing

\abstract
\noindent
In this paper we introduce the class of beta seasonal autoregressive moving average ($\beta$SARMA) models for modeling and forecasting time series data that assume values in the standard unit interval. It generalizes the class of beta autoregressive moving average models [Rocha and Cribari-Neto, Test, 2009] by incorporating seasonal dynamics to the model dynamic structure. Besides introducing the new class of models, we develop parameter estimation, hypothesis testing inference, and diagnostic analysis tools. We also discuss out-of-sample forecasting. In particular, we provide closed-form expressions for the conditional score vector and for the conditional Fisher information matrix. We also evaluate the finite sample performances of conditional maximum likelihood estimators and white noise tests using Monte Carlo simulations. An empirical application is presented and discussed.




\section{Introduction}
\label{S:introduction}

Univariate time series modeling is commonly used in many fields. Most conventional time series models are based on the Gaussianity assumption~\cite{Chuang2007}. A well-known class of this linear models is the class of autoregressive integrated moving average models  (ARIMA)~\cite{Box2008}. However, it has been recognized that the Gaussian assumption is too restrictive for many applications~\cite{Tiku2000}.
As a consequence, there has been increased interest in non-Gaussian time series models~\cite{Zheng2015}.
Some models for discrete variate time series are considered in \cite{McKenzie1985,Osh1987,Ristic2012}.
In \cite{Zeger1988} is proposed a quasi-likelihood approach to regression models for discrete and continuous time series.
In \cite{Li1988} is focused on time series modeling under non-Gaussian innovations.
Non-Gaussian time series models are considered as instantaneous transformations of Gaussian time series in \cite{Janacek1990, Chuang2007}.
Time series models based on generalized linear models (GLM)~\cite{McCullagh1989} are considered in \cite{Li1991,Li1994,Benjamin2003,Fokianos2004}.
Other important and recent works on non-Gaussian time series modeling are \cite{Sim1994,Swift1995,Tiku2000,Akkaya2005,Jung2006,Bandon2009,Zheng2015}.
A comprehensive reference on general models for time series analysis is \cite{Kedem2002}.

Practitioners are oftentimes interested in modeling the behavior of variables that assume values in the standard unit interval, $(0,1)$, such as rates and proportions~\cite{Kumaraswamy1980, Grunwald1993, Ferrari2004, Johnson1995, Kieschnick2003, Melo2009, Gupta2011, Guolo2014}.
Time series modeling of such variables can be accomplished by using the class of beta autoregressive moving average models ($\beta$ARMA)~\cite{Rocha2009}. It is noteworthy that the beta distribution is quite flexible since its density can assume a variety of different shapes depending on the values of the parameters that index the distribution: it can be symmetric, left-skewed, right-skewed, constant, J-shaped, and inverted J-shaped~\cite{Ferrari2004, Rocha2009,Bonat2015}.
According to \cite{Bonat2015}, ``there are situations where the response variable is continuous and bounded above and below such as rates, percentages, indexes and proportions. In such situations, the traditional GLMM based on the Gaussian distribution is not adequate, since bounding is ignored. An approach that has been used to model this type of data is based on the beta distribution.'' Recent related works include \cite{Casarin2012,Guolo2014,daSilva2011,Ferreira2015}.

Time series data may exhibit periodical fluctuations, i.e., they may display seasonality. Models that include seasonality have been extensively explored in the literature~\cite{Lund2000, Basawa2004, Monteiro2010}, the seasonal ARIMA model (SARIMA)~\cite{Box2008} being the most used model for Gaussian seasonal data. A commonly used approach for dealing with non-Gaussian data is to assume that seasonal fluctuations are deterministic and then model them using sine/cosine functions as covariates in regression times series models; see \cite{Guolo2014,Morina2011,Benjamin2003}. Such an strategy, however, is not appropriate when the seasonality is driven by a stochastic mechanism~\cite{Briet2013}. Some authors have recently devoted attention to non-Gaussian seasonal time series models; see, e.g., \cite{Briet2013,Bourguignon2015}. To the best of our knowledge, however, no seasonal time series model is available for variables that assume values in the standard unit interval, such as rates and proportions.

Our chief goal is to introduce a time series model based on the beta law that includes stochastic seasonal dynamics: the beta seasonal autoregressive moving average model ($\beta$SARMA). We also outline maximum likelihood parameter estimation,
obtain closed-form expressions for the conditional score function and for the conditional Fisher information matrix,
show how confidence intervals can be constructed and how hypothesis testing inference can be performed (including a seasonality test), address the issue of model selection, propose different residuals that can be used to assess goodness-of-fit, present white noise tests based on such residuals, and show how out-of-sample forecasts can be produced.
Additionally,
we present results from Monte Carlo simulations that were carried out to evaluate the accuracy of maximum likelihood estimation and white noise testing inference in finite samples.
Finally, we present and discuss an empirical application. We note that the proposed $\beta$SARMA finds potential applications in a plethora of scientific areas, such as
mortality rate~\cite{Mitchell2013},
seasonal infectious disease~\cite{Dietz1976,Dowel2001,Grassly2006},
unemployment rate~\cite{Ferreira2015, Rocha2009},
seasonal variation of fixed and volatile oil percentage~\cite{Emara2011},
solar radiation~\cite{Mullen2012},
periodic ocean waves~\cite{Sundar1989},
and also in hydrological applications~\cite{Fletcher1996,Ganji2006}.

The paper unfolds as follows.
Section~\ref{s:modelo} introduces the seasonal beta autoregressive moving average model.
Several particular cases of the proposed model are examined. Parameter estimation via conditional maximum likelihood is outlined in Section~\ref{S:estimation}.
We provide closed-form expressions for the first derivatives of conditional log-likelihood function (score function) and for the conditional information matrix. Interval estimation and hypothesis testing strategies are also presented.
Section~\ref{S:diagnostic} addresses model selection, residuals, and diagnostic analysis.
Section~\ref{S:simulation} contains Monte Carlo simulation results on parameter estimation and white noise testing.
An empirical application is presented and discussed in Section~\ref{S:application}.
Finally, Section~\ref{S:conclusions} offers some concluding remarks.

\section{The proposed model}\label{s:modelo}

\begin{figure}[t]
\begin{center}
\subfigure[$\;\varphi=20$]
{\label{f:densitys_10}\includegraphics[width=0.4\textwidth]{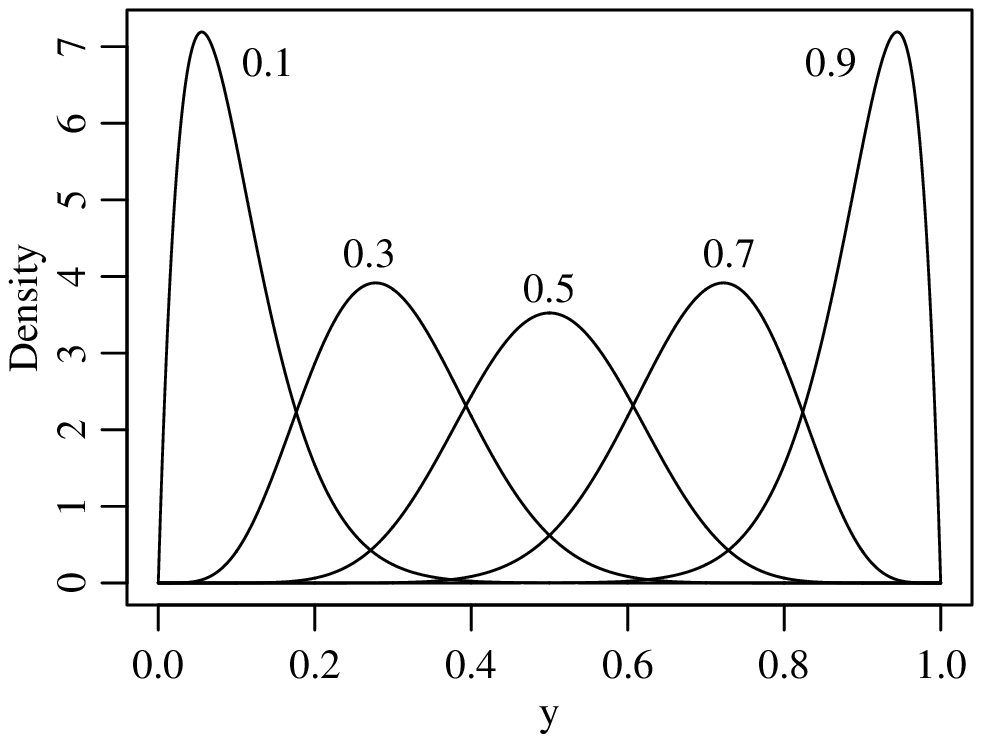}}
\subfigure[$\;\varphi=120$]
{\label{f:densitys_90}\includegraphics[width=0.4\textwidth] {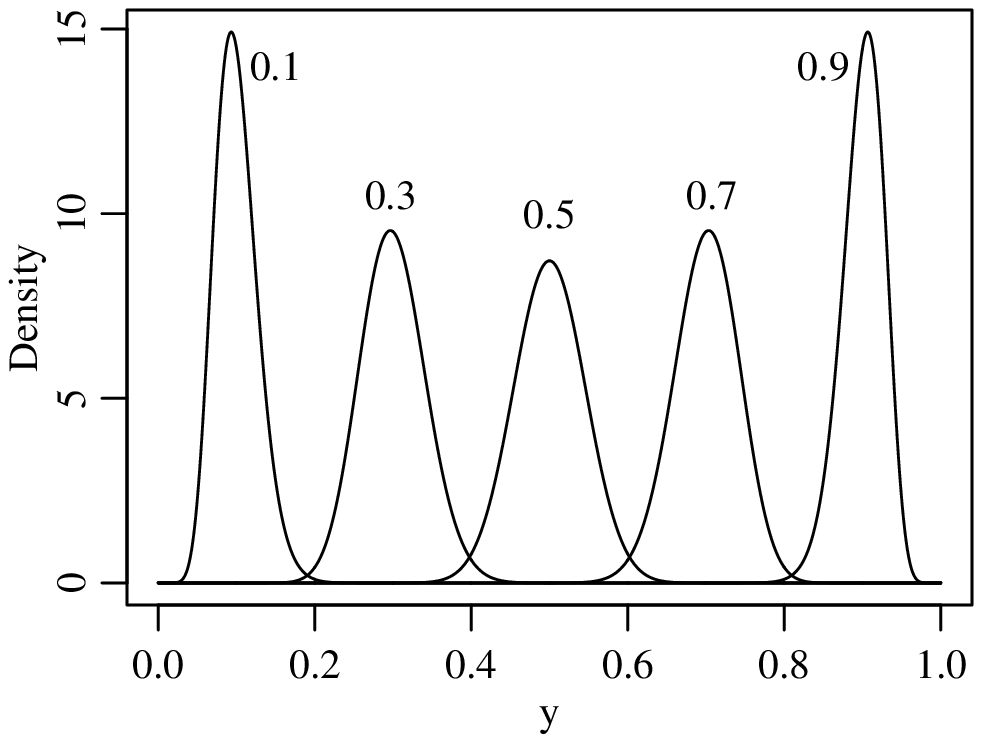}}
\caption{
Beta density functions:
$\mu=0.10$, $0.30$, $0.50$, $0.70$, $0.90$ and two values of $\varphi$.
}\label{f:density}
\end{center}
\end{figure}

Let $\boldsymbol{y} = (y_1,\ldots,y_n)^\top$ be a vector of $n$ random variables, each $y_t$, $t=1,2,\ldots,n$, being beta distributed conditional on the set of previous information $\mathcal{F}_{t-1}$. The distribution parameters are  $\mu_t$ (mean) and $\varphi$ (precision). The conditional density of $y_t$ given $\mathcal{F}_{t-1}$ is
\begin{align*}
f(y_t\mid\mathcal{F}_{t-1})
=
\frac{\Gamma(\varphi)}{\Gamma(\mu_t\varphi)\Gamma((1-\mu_t)\varphi)}
y_t^{\mu_t\varphi-1}
(1-y_t)^{(1-\mu_t)\varphi-1}
,
\quad
0<y_t<1
,
\end{align*}
where $0<\mu_t<1$ and $\varphi>0$. Figure~\ref{f:density} presents beta densities for different parameter values. The conditional mean and the conditional variance of $y_t$ are given by
\begin{align*}
\E(y_t\mid \mathcal{F}_{t-1}) & = \mu_t
,
\\
{\rm Var}(y_t \mid \mathcal{F}_{t-1}) &= V(\mu_t)/(1+\varphi)
,
\end{align*}
respectively, where $V(\mu_t)=\mu_t(1-\mu_t)$ is the variance function and $\varphi$ can be interpreted as a precision parameter (reciprocal of dispersion).

Rocha and Cribari-Neto \cite{Rocha2009} introduced a dynamic model that can be used to model the behavior of variables that assume values in the standard unit interval. Their model, however, cannot be used when the variable of interest is subject to stochastic seasonal fluctuations. We shall now extend their model so that it can be used when such fluctuations do exist. Our interest lies in modeling the conditional mean of $y_t$.

The proposed beta seasonal autoregressive moving average model, $\beta$SARMA$(p,q)\times (P,Q)_S$, is given by
\begin{align}\label{E:bsarma}
\Phi(B^S) \phi(B)g(y_t) = \beta + \Theta(B^S) \theta(B) r_t,
\end{align}
where
$\beta \in \R$ is a constant, $r_t=g(y_t)-g(\mu_t)$ is the error term,
$g(\cdot)$ is a strictly monotone and twice differentiable link function such that $g:(0,1)\rightarrow \mathbb{R} $,
$\phi(B) = 1 - \phi_1 B - \phi_2 B^2 - \cdots - \phi_p B^p$ is the autoregressive polynomial of order $p$,
$\theta(B) = 1 - \theta_1 B - \theta_2 B^2 - \cdots - \theta_q B^q$ is the moving average polynomial of order $q$,
$\Phi(B^S) = 1 - \Phi_1 B^S - \Phi_2 B^{2S} - \cdots - \Phi_P B^{PS}$ is the seasonal autoregressive polynomial of order $P$,
$\Theta(B^S) = 1 - \Theta_1 B^S - \Theta_2 B^{2S} - \cdots - \Theta_Q B^{QS}$ is the seasonal moving average polynomial of order $Q$,
$B$ is the backshift operator such that
$B^d g(y_t) = g(y_{t-d})$,
and
$B^d r_t = r_{t-d}$ for a nonnegative integer~$d$;
and
$S$ is the seasonality frequency
(typically, $S=12$ for monthly data and $S=4$ for quarterly data).

\subsection{Some particular cases}

The class of $\beta$SARMA$(p,q)\times (P,Q)_S$ contains several important models as particular cases. Some of them are listed below.

\subsubsection{$\beta$ARMA$(1,1)$ or $\beta$SARMA$(1,1)\times (0,0)_S$ }

The $\beta$ARMA$(1,1)$ can be written as
\begin{align*}
&\phi(B)g(y_t) = \beta + \theta(B) r_t \\
&(1 - \phi_1 B)g(y_t) = \beta + (1 - \theta_1 B) r_t \\
&g(y_t) - \phi_1 g(y_{t-1}) = \beta + r_t - \theta_1 r_{t-1} \\
&g(y_t) - \phi_1 g(y_{t-1}) = \beta + [g(y_t)-g(\mu_t)] - \theta_1 r_{t-1} \\
&g(\mu_t) = \beta + \phi_1 g(y_{t-1}) - \theta_1 r_{t-1}= \eta_t,
\end{align*}
where $g(\mu_t)=\eta_t$ is the linear predictor.

\subsubsection{$\beta$ARMA$(p,q)$}

Using backshift operator the \barma$(p,q)$ model %
can be written as
\begin{align*}%
&\phi(B)g(y_t) = \beta + \theta(B) r_t\\
&(1 - \phi_1 B - \cdots - \phi_p B^p)g(y_t)
= \beta + (1 - \theta_1 B - \cdots - \theta_q B^q) r_t \\
&g(y_t) - \phi_1 g(y_{t-1}) - \cdots - \phi_p g(y_{t-p})
= \beta + r_t - \theta_1 r_{t-1}- \cdots - \theta_q r_{t-q} \\
&g(y_t) - \phi_1 g(y_{t-1}) - \cdots - \phi_p g(y_{t-p})
= \beta + [g(y_t)-g(\mu_t)] - \theta_1 r_{t-1}- \cdots - \theta_q r_{t-q} \\
&g(\mu_t) = \beta + \sum_{i=1}^p \phi_i g(y_{t-i}) - \sum_{j=1}^{q}\theta_j r_{t-j}.
\end{align*}
Such a model is thus a special case of the more general class of models we propose in this paper.

\subsubsection{$\beta$SARMA$(1,1)\times (1,1)_{12}$}

The $\beta$SARMA$(1,1)\times (1,1)_{12}$ model can be written as
\begin{align*}
& \Phi(B^{12}) \phi(B)g(y_t) = \beta + \Theta(B^{12}) \theta(B) r_t \\
& (1- \Phi_1 B^{12})(1- \phi_1 B)g(y_t) = \beta + (1- \Theta_1 B^{12})(1- \theta_1 B) r_t \\
&g(y_t) - \phi_1 g(y_{t-1}) - \Phi_1 g(y_{t-12}) + \phi_1 \Phi_1 g(y_{t-13})
=
\beta + r_t - \theta_1 r_{t-1} - \Theta_1 r_{t-12} + \theta_1 \Theta_1 r_{t-13}  \\
&g(\mu_t) = \beta + \phi_1 g(y_{t-1}) + \Phi_1 g(y_{t-12}) - \phi_1 \Phi_1 g(y_{t-13}) - \theta_1 r_{t-1} - \Theta_1 r_{t-12} + \theta_1 \Theta_1 r_{t-13}.
\end{align*}

\subsubsection{$\beta$SARMA$(2,0)\times (2,0)_{12}$}

The $\beta$SARMA$(2,0)\times (2,0)_{12}$ model is given by
\begin{align*}
& \Phi(B^{12}) \phi(B)g(y_t) = \beta + r_t \\
& (1- \Phi_1 B^{12} - \Phi_2 B^{24})(1- \phi_1 B - \phi_2 B^2)g(y_t) = \beta + r_t \\
&g(y_t) - \phi_1 g(y_{t-1}) - \phi_2 g(y_{t-2}) - \Phi_1 g(y_{t-12}) - \Phi_2 g(y_{t-24}) + \phi_1 \Phi_1 g(y_{t-13})\\
& \qquad + \phi_2 \Phi_1 g(y_{t-14}) + \phi_1 \Phi_2 g(y_{t-25}) + \phi_2 \Phi_2 g(y_{t-26})
= \beta + r_t  \\
&g(\mu_t) = \beta + \phi_1 g(y_{t-1}) + \phi_2 g(y_{t-2}) + \Phi_1 g(y_{t-12}) + \Phi_2 g(y_{t-24}) - \phi_1 \Phi_1 g(y_{t-13}) - \phi_2 \Phi_1 g(y_{t-14}) \\
&\qquad - \phi_1 \Phi_2 g(y_{t-25})
- \phi_2 \Phi_2 g(y_{t-26}).
\end{align*}

\section{Parameter estimation}\label{S:estimation}

Parameter estimation can be carried out by conditional maximum likelihood~\cite{Andersen1970}.
Let $\boldsymbol{\gamma}=(\beta,\boldsymbol{\phi}^\top,\boldsymbol{\theta}^\top,\boldsymbol{\Phi}^\top,\boldsymbol{\Theta}^\top,\varphi)^\top$ be the $k$-dimensional parameter vector, where
$\boldsymbol{\phi}=(\phi_1,\ldots,\phi_p)^\top$,
$\boldsymbol{\theta}=(\theta_1,\ldots,\theta_q)^\top$,
$\boldsymbol{\Phi}=(\Phi_1,\ldots,\Phi_P)^\top$,
$\boldsymbol{\Theta}=(\Theta_1,\ldots,\Theta_Q)^\top$,
and $k=p+q+P+Q+2$. The conditional maximum likelihood estimators (CMLE) of $\gamma$ are obtained by maximizing the logarithm of the conditional likelihood function.
The log-likelihood function for the parameter vector $\gamma$ conditional on the $m$ initial observations, where $m=\max(PS+p, QS+q)$,
can be written as
\begin{align}\label{E:loglik}
\ell=\ell(\boldsymbol{\gamma};\boldsymbol{y})=\sum\limits_{t=m+1}^{n} \log f(y_t\mid\mathcal{F}_{t-1})= \sum\limits_{t=m+1}^{n}\ell_t(\mu_t,\varphi),
\end{align}
where $\ell_t(\mu_t,\varphi)=\log\Gamma(\varphi)-\log\Gamma(\mu_t\varphi)-\log\Gamma((1-\mu_t)\varphi) + (\mu_t \varphi-1)\log y_t+\lbrace (1-\mu_t)\varphi -1\rbrace \log (1-y_t)$.

\subsection{Conditional score vector}\label{s:score}

Let $\boldsymbol{\lambda}=(\beta,\boldsymbol{\phi}^\top,\boldsymbol{\theta}^\top,\boldsymbol{\Phi}^\top,\boldsymbol{\Theta}^\top)^\top$. Differentiation of the conditional log-likelihood function given in \eqref{E:loglik} with respect to
$i$th element of $\boldsymbol{\lambda}$, $\lambda_i$, with $i=1,\ldots,(k-1)$,
yields
\begin{align*}
 \frac{\partial \ell}{\partial \lambda_i} &=
 \sum\limits_{t=m+1}^{n} \frac{\partial \ell_t(\mu_t,\varphi)}{\partial \mu_t}
 \frac{d \mu_t}{d \eta_t} \frac{\partial \eta_t}{\partial \lambda_i}.
\end{align*}
Since
\begin{align}\label{e:dldmu}
\frac{\partial \ell_t(\mu_t,\varphi)}{\partial \mu_t}
= \varphi
\left\{
\log{\frac{y_t}{1-y_t}} -
\left[
\psi (\mu_t \varphi) - \psi
\left(
(1-\mu_t)\varphi
\right)
\right]
\right\},
\end{align}
where $\psi(\cdot)$ is the digamma function,
it follows that
\begin{align*}
 \frac{\partial \ell}{\partial \lambda_i} &= \varphi \sum \limits_{t=m+1}^{n}(y_t^\ast-\mu_t^\ast)\frac {1}{g^\shortmid(\mu_t)}\frac{\partial \eta_t}{\partial \lambda_i},
\end{align*}
where $y_t^ {\ast} = \log\lbrace y_t / (1-y_t)\rbrace$,
$\mu_t^ {\ast} = \psi(\mu_t\varphi)-\psi((1-\mu_t)\varphi)$.

When $\lambda_i = \beta$, the linear predictor derivative is
\begin{align*}
\frac{\partial \eta_t}{\partial \beta}= 1
- \sum_{j=1}^{q}\theta_j \frac{\partial r_{t-j}}{\partial \beta}
- \sum_{J=1}^{Q}\Theta_J \frac{\partial r_{t-JS}}{\partial \beta}
+ \sum_{j=1}^{q}\sum_{J=1}^{Q} \theta_j\Theta_J \frac{\partial r_{t-(j+JS)}}{\partial \beta}.
\end{align*}
Since that $r_t= g(y_t) - \eta_t$, we obtain
\begin{align*}
\frac{\partial \eta_t}{\partial \beta}= 1
+ \sum_{j=1}^{q}\theta_j \frac{\partial \eta_{t-j}}{\partial \beta}
+ \sum_{J=1}^{Q}\Theta_J \frac{\partial \eta_{t-JS}}{\partial \beta}
- \sum_{j=1}^{q}\sum_{J=1}^{Q} \theta_j\Theta_J \frac{\partial \eta_{t-(j+JS)}}{\partial \beta}.
\end{align*}

The linear predictor derivatives with respect to the remaining parameters are given by
\begin{align*}
\frac{\partial \eta_t}{\partial \phi_i} = g(y_{t-i})\Phi(B^S)
 + \sum_{j=1}^{q}\theta_j \frac{\partial \eta_{t-j}}{\partial \phi_i}
+ \sum_{J=1}^{Q}\Theta_J \frac{\partial \eta_{t-JS}}{\partial \phi_i}
- \sum_{j=1}^{q}\sum_{J=1}^{Q} \theta_j\Theta_J \frac{\partial \eta_{t-(j+JS)}}{\partial \phi_i},
\\
 \frac{\partial \eta_t}{\partial \Phi_I} = g(y_{t-IS})\phi(B)
 + \sum_{j=1}^{q}\theta_j \frac{\partial \eta_{t-j}}{\partial \Phi_I}
+ \sum_{J=1}^{Q}\Theta_J \frac{\partial \eta_{t-JS}}{\partial \Phi_I}
- \sum_{j=1}^{q}\sum_{J=1}^{Q} \theta_j\Theta_J \frac{\partial \eta_{t-(j+JS)}}{\partial \Phi_I},
\\
 \frac{\partial \eta_t}{\partial \theta_j} = -r_{t-j}\Theta(B^S)
  + \sum_{i=1}^{q}\theta_i \frac{\partial \eta_{t-i}}{\partial \theta_j}
+ \sum_{J=1}^{Q}\Theta_J \frac{\partial \eta_{t-JS}}{\partial \theta_j}
- \sum_{i=1}^{q}\sum_{J=1}^{Q} \theta_i\Theta_J \frac{\partial \eta_{t-(i+JS)}}{\partial \theta_j}, \\
 \frac{\partial \eta_t}{\partial \Theta_J} = -r_{t-JS}\theta(B)
+ \sum_{j=1}^{q}\theta_j \frac{\partial \eta_{t-j}}{\partial \Theta_J}
+ \sum_{i=1}^{Q}\Theta_i \frac{\partial \eta_{t-iS}}{\partial \Theta_J}
- \sum_{j=1}^{q}\sum_{i=1}^{Q} \theta_j\Theta_i \frac{\partial \eta_{t-(j+iS)}}{\partial \Theta_J}.
\end{align*}

As in \cite{Benjamin1998} and \cite{Rocha2017}, when there are no moving average components (i.e., when $\theta_j=0$ and $\Theta_J=0$ for all $j$ and $J$) no recursion is necessary to evaluate $\eta_t$ and its partial derivatives. When the model includes moving average dynamics (ordinary or seasonal), we suggest using $\eta_t = g(y_t)$ and setting all linear predictor derivatives equal to zero for the initial cases, as in \cite{Rocha2017}.

Finally, differentiation of $\ell(\boldsymbol{\gamma};\boldsymbol{y})$
with respect to the precision parameter $\varphi$ yields
\begin{align}\label{e:dldvarphi}
\frac{\partial \ell}{\partial \varphi} = \sum \limits _ {t=m+1} ^ {n} \lbrace \mu_t (y_t^\ast - \mu_t^\ast) + \log (1-y_t) - \psi ((1-\mu_t)\varphi)+\psi(\varphi)\rbrace.
\end{align}

Therefore, the elements of the score vector
$\boldsymbol{U}(\boldsymbol{\gamma})=\left(
U_{\beta}(\boldsymbol{\gamma}),
\boldsymbol{U}_{\phi}(\boldsymbol{\gamma})^{\top},
\boldsymbol{U}_{\theta}(\boldsymbol{\gamma})^{\top},
\boldsymbol{U}_{\Phi}(\boldsymbol{\gamma})^{\top},
\boldsymbol{U}_{\Theta}(\boldsymbol{\gamma})^{\top},
U_{\varphi}(\boldsymbol{\gamma})\right)^{\top}$
can be written, in matrix form, as
\begin{align*}
U_\beta (\boldsymbol{\gamma}) &= \varphi \boldsymbol{a}^\top \boldsymbol{T} (\boldsymbol{y}^\ast - \boldsymbol{\mu}^\ast), \\
\boldsymbol{U}_\phi (\boldsymbol{\gamma}) &= \varphi \boldsymbol{A}^\top \boldsymbol{T}(\boldsymbol{y}^\ast - \boldsymbol{\mu}^\ast), \\
\boldsymbol{U}_\Phi (\boldsymbol{\gamma}) &= \varphi \mathbfcal{A}^\top \boldsymbol{T}(\boldsymbol{y}^\ast - \boldsymbol{\mu}^\ast), \\
\boldsymbol{U}_\theta (\boldsymbol{\gamma}) &= \varphi \boldsymbol{M}^\top \boldsymbol{T}(\boldsymbol{y}^\ast - \boldsymbol{\mu}^\ast), \\
\boldsymbol{U}_\Theta (\boldsymbol{\gamma}) &= \varphi \mathbfcal{M}^\top \boldsymbol{T}(\boldsymbol{y}^\ast - \boldsymbol{\mu}^\ast), \\
U_\varphi (\boldsymbol{\gamma}) &= \sum \limits _ {t=m+1} ^ {n} \lbrace \mu_t (y_t^\ast - \mu_t^\ast) + \log (1-y_t) - \psi ((1-\mu_t)\varphi)+\psi(\varphi)\rbrace,
\end{align*}
where $\boldsymbol{y}^\ast = (y_{m+1}^\ast,\ldots,y_n^\ast)^\top$,
$\boldsymbol{\mu}^\ast=(\mu_{m+1}^\ast,\ldots,\mu_n^\ast)^\top$,
$\boldsymbol{T}= \mathrm{diag}\lbrace 1/g^\shortmid(\mu_{m+1}) , \ldots ,1/g^\shortmid(\mu_n)\rbrace$,
$\boldsymbol{a} = \left(\frac{\partial \eta_{m+1}}{\partial \beta},\ldots,
\frac{\partial \eta_{n}}{\partial \beta} \right)^\top$,
$\boldsymbol{A}$ is an $(n-m)\times p$ matrix whose $(i,j)$ element is given by
${\partial \eta_{i+m}}/{\partial \phi_j}$,
$\mathbfcal{A}$ is an $(n-m)\times P$ matrix whose $(i,j)$ element equals
${\partial \eta_{i+m}}/{\partial \Phi_j}$,
$\boldsymbol{M}$ is an $(n-m)\times q$ matrix whose $(i,j)$ element is given by
${\partial \eta_{i+m}}/{\partial \theta_j}$,
and
$\mathbfcal{M}$ is an $(n-m)\times Q$ matrix whose $(i,j)$ element is
${\partial \eta_{i+m}}/{\partial \Theta_j}$,

The conditional maximum likelihood estimator of $\boldsymbol{\gamma}$ is obtained as the solution to the following system of equations:
\begin{align*}
\boldsymbol{U}(\boldsymbol{\gamma})=\boldsymbol{0},
\end{align*}
where $\boldsymbol{0}$ is the $k\times1$ vector of zeros.
The solution to the above system of nonlinear equation cannot be written in closed form. Conditional maximum likelihood estimates can be  obtained by numerically maximizing the conditional log-likelihood function using a Newton or quasi-Newton nonlinear optimization algorithm; see, e.g., \cite{nocedal1999}. In what follows, we shall use the quasi-Newton algorithm known as
Broyden-Fletcher-Goldfarb-Shanno (BFGS); for details, see \cite{press}.

\subsection{Conditional information matrix}

The CMLE asymptotic covariance matrix, which can be used to construct confidence intervals, is given by the inverse of the conditional Fisher information matrix. In order to obtain such a matrix we need to compute the expected values of all second order derivatives.

Let $\boldsymbol{\lambda}=(\beta,\boldsymbol{\phi}^\top,\boldsymbol{\theta}^\top,\boldsymbol{\Phi}^\top,\boldsymbol{\Theta}^\top)^\top$. It can be shown that
\begin{align*}
\frac{\partial^2\ell}{\partial \lambda_i \partial \lambda_j} &= \sum_{t=m+1}^{n}\frac{\partial}{\partial \mu_t}
\left( \frac{\partial \ell_t(\mu_t,\varphi)}{\partial \mu_t}\frac{d \mu_t}{d \eta_t} \frac{\partial \eta_t}{\partial \lambda_j}\right)
\frac{d \mu_t}{d \eta_t} \frac{\partial \eta_t}{\partial \lambda_i} \\
&= \sum_{t=m+1}^{n} \left[ \frac{\partial^2 \ell_t(\mu_t,\varphi)}{\partial \mu_t^2}\frac{d \mu_t}{d \eta_t} \frac{\partial \eta_t}{\partial \lambda_j}
+ \frac{\partial \ell_t(\mu_t,\varphi)}{\partial \mu_t}\frac{\partial}{\partial \mu_t}\left(\frac{d \mu_t}{d \eta_t} \frac{\partial \eta_t}{\partial \lambda_j} \right) \right]
\frac{d \mu_t}{d \eta_t} \frac{\partial \eta_t}{\partial \lambda_i},
\end{align*}
for $i=1, \ldots,(k-1)$ and $j=1, \ldots,(k-1)$, where $k=p+q+P+Q+2$.

Under the usual regularity conditions, it follows that $\E(\partial \ell_t(\mu_t,\varphi)/\partial \mu_t | \mathcal{F}_{t-1})=0$. Thus,
\begin{align*}
\E\left( \left. \frac{\partial^2\ell}{\partial \lambda_i \partial \lambda_j}  \right| \mathcal{F}_{t-1} \right)
&= \sum_{t=m+1}^{n} \E\left( \left. \frac{\partial^2 \ell_t(\mu_t,\varphi)}{\partial \mu_t^2} \right| \mathcal{F}_{t-1} \right)
\left(\frac{d \mu_t}{d \eta_t} \right)^2
\frac{\partial \eta_t}{\partial \lambda_j}
\frac{\partial \eta_t}{\partial \lambda_i}.
\end{align*}
By differentiating \eqref{e:dldmu} twice with respect to $\mu_t$, we obtain
\begin{align*}
\frac{\partial^2 \ell_t(\mu_t,\varphi)}{\partial \mu_t^2} = -\varphi^2
\left\{
\psi' (\mu_t \varphi) +
\psi'
\left[
(1-\mu_t)\varphi
\right]
\right\}.
\end{align*}
Furthermore,
\begin{align*}
\E\left( \left. \frac{\partial^2\ell}{\partial \lambda_i \partial \lambda_j}  \right| \mathcal{F}_{t-1} \right)
&= - \sum_{t=m+1}^{n}
\frac{w_t}{g'(\mu_t)^2}
\frac{\partial \eta_t}{\partial \lambda_j}
\frac{\partial \eta_t}{\partial \lambda_i},
\end{align*}
where
$w_t=\varphi^2 \left\{
\psi' (\mu_t \varphi) +
\psi'
\left[
(1-\mu_t)\varphi
\right]
\right\} $.
Notice that all first derivatives $\partial \eta_t/\partial \lambda_i$ have already been presented in Section~\ref{s:score}.

Differentiation of \eqref{e:dldvarphi} with respect to $\lambda_i$, $i=1,\ldots,(k-1)$, yields
\begin{align*}
\frac{\partial^2 \ell}{\partial \varphi \partial \lambda_i} = \sum \limits _ {t=m+1} ^ {n}
\left[ (y_t^\ast - \mu_t^\ast) - \varphi \frac{\partial \mu_t^\ast}{\partial \varphi} \right]
\frac{1}{g'(\mu_t)}\frac{\partial \eta_t}{\partial \lambda_i},
\end{align*}
where ${\partial \mu_t^\ast}/{\partial \varphi} =
\psi' (\mu_t \varphi) \mu_t
-\psi'
\left[
(1-\mu_t)\varphi
\right] (1-\mu_t)$.
Under the usual regularity conditions, we have $\E(y^\ast_t | \mathcal{F}_{t-1})=\mu_t^\ast$, and thus
\begin{align*}
\E\left( \left. \frac{\partial^2 \ell}{\partial \varphi \partial \lambda_i}  \right| \mathcal{F}_{t-1} \right)
&= - \sum_{t=m+1}^{n}
\frac{c_t}{g'(\mu_t)}
\frac{\partial \eta_t}{\partial \lambda_i},
\end{align*}
where
$ c_t = \varphi \frac{\partial \mu_t^\ast}{\partial \varphi} = \varphi \left\{\psi' (\mu_t \varphi) \mu_t
-\psi'
\left[
(1-\mu_t)\varphi
\right] (1-\mu_t)\right\}$.

Finally, the expected value of the second order derivative of $\ell(\gamma;y)$ with respect to $\varphi$ is given by
\begin{align*}
\E\left( \left. \frac{\partial^2 \ell}{\partial \varphi^2}  \right| \mathcal{F}_{t-1} \right) =
- \sum \limits _ {t=m+1} ^ {n} d_t,
\end{align*}
where $d_t=\psi' (\mu_t\phi)\mu_t^2+\psi'((1-\mu_t)\phi)(1-\mu_t)^2-\psi'(\phi)$.

Let
$\boldsymbol{W}=\mathrm{diag}\{w_{m+1},\ldots,w_n\}$,
$\boldsymbol{C}=\mathrm{diag}\{c_{m+1},\ldots,c_n\}$,
and
$\boldsymbol{D} = \mathrm{diag}\{d_{m+1},\ldots,d_n\}$.
The joint conditional Fisher information matrix for $\boldsymbol{\gamma}$ is
\begin{align*}
\boldsymbol{K}=
\boldsymbol{K}(\boldsymbol{\gamma}) = \left( \begin{array}{cccccc}
{K}_{(\beta,\beta)} &\boldsymbol{K}_{(\beta,\phi)} &\boldsymbol{K}_{(\beta,\Phi)} &\boldsymbol{K}_{(\beta,\theta)} &\boldsymbol{K}_{(\beta,\Theta)} &\boldsymbol{K}_{(\beta,\varphi)} \\
\boldsymbol{K}_{(\phi,\beta)} &\boldsymbol{K}_{(\phi,\phi)} &\boldsymbol{K}_{(\phi,\Phi)}  &\boldsymbol{K}_{(\phi,\theta)} &\boldsymbol{K}_{(\phi,\Theta)} &\boldsymbol{K}_{(\phi,\varphi)}\\
\boldsymbol{K}_{(\Phi,\beta)} &\boldsymbol{K}_{(\Phi,\phi)} &\boldsymbol{K}_{(\Phi,\Phi)}  &\boldsymbol{K}_{(\Phi,\theta)} &\boldsymbol{K}_{(\Phi,\Theta)} &\boldsymbol{K}_{(\Phi,\varphi)}\\
\boldsymbol{K}_{(\theta,\beta)} &\boldsymbol{K}_{(\theta,\phi)} &\boldsymbol{K}_{(\theta,\Phi)}  &\boldsymbol{K}_{(\theta,\theta)} &\boldsymbol{K}_{(\theta,\Theta)} &\boldsymbol{K}_{(\theta,\varphi)} \\
\boldsymbol{K}_{(\Theta,\beta)} &\boldsymbol{K}_{(\Theta,\phi)} &\boldsymbol{K}_{(\Theta,\Phi)}  &\boldsymbol{K}_{(\Theta,\theta)} &\boldsymbol{K}_{(\Theta,\Theta)} &\boldsymbol{K}_{(\Theta,\varphi)} \\
\boldsymbol{K}_{(\varphi,\beta)} &\boldsymbol{K}_{(\varphi,\phi)} &\boldsymbol{K}_{(\varphi,\Phi)}  &\boldsymbol{K}_{(\varphi,\theta)} &\boldsymbol{K}_{(\varphi,\Theta)} & {K}_{(\varphi,\varphi)}
\end{array} \right),
\end{align*}
where
${K}_{(\beta,\beta)} = \boldsymbol{a}^\top \boldsymbol{W}\boldsymbol{T}^2 \boldsymbol{a}$, %
$\boldsymbol{K}_{(\beta,\phi)} = \boldsymbol{K}^\top_{(\phi,\beta)} = \boldsymbol{a}^\top \boldsymbol{W} \boldsymbol{T}^2 \boldsymbol{A}$, %
$\boldsymbol{K}_{(\beta,\Phi)} = \boldsymbol{K}^\top_{(\Phi,\beta)} = \boldsymbol{a}^\top \boldsymbol{W} \boldsymbol{T}^2 \mathbfcal{A}$, %
$\boldsymbol{K}_{(\beta,\theta)} = \boldsymbol{K}_{(\theta,\beta)}^\top = \boldsymbol{a}^\top \boldsymbol{W} \boldsymbol{T}^2 \boldsymbol{M}$, %
$\boldsymbol{K}_{(\beta,\Theta)} = \boldsymbol{K}_{(\Theta,\beta)}^\top = \boldsymbol{a}^\top \boldsymbol{W} \boldsymbol{T}^2 \mathbfcal{M}$, %
$\boldsymbol{K}_{(\beta,\varphi)} = \boldsymbol{K}_{(\varphi,\beta)} = \boldsymbol{a}^\top \boldsymbol{C}\boldsymbol{T} \textbf{1}$,
$\boldsymbol{K}_{(\phi,\phi)} = \boldsymbol{A}^\top \boldsymbol{W} \boldsymbol{T}^2 \boldsymbol{A}$, %
$\boldsymbol{K}_{(\phi,\Phi)} = \boldsymbol{K}_{(\Phi,\phi)}^\top = \boldsymbol{A}^\top \boldsymbol{W} \boldsymbol{T}^2\mathbfcal{A}$, %
$\boldsymbol{K}_{(\phi,\theta)} = \boldsymbol{K}_{(\theta,\phi)}^\top = \boldsymbol{A}^\top \boldsymbol{W} \boldsymbol{T}^2 \boldsymbol{M}$, %
$\boldsymbol{K}_{(\phi,\Theta)} = \boldsymbol{K}_{(\Theta,\phi)}^\top = \boldsymbol{A}^\top \boldsymbol{W} \boldsymbol{T}^2 \mathbfcal{M}$, %
$\boldsymbol{K}_{(\phi,\varphi)} = \boldsymbol{K}_{(\varphi,\phi)}^\top = \boldsymbol{A}^\top \boldsymbol{C}\boldsymbol{T} \textbf{1}$, %
$\boldsymbol{K}_{(\Phi,\Phi)} = \mathbfcal{A}^\top \boldsymbol{W} \boldsymbol{T}^2 \mathbfcal{A}$, %
$\boldsymbol{K}_{(\Phi,\theta)} = \boldsymbol{K}_{(\theta,\Phi)}^\top = \mathbfcal{A}^\top \boldsymbol{W} \boldsymbol{T}^2 \boldsymbol{M}$, %
$\boldsymbol{K}_{(\Phi,\Theta)} = \boldsymbol{K}_{(\Theta,\Phi)}^\top = \mathbfcal{A}^\top \boldsymbol{W} \boldsymbol{T}^2 \mathbfcal{M}$, %
$\boldsymbol{K}_{(\Phi,\varphi)} = \boldsymbol{K}_{(\varphi,\Phi)}^\top = \mathbfcal{A}^\top \boldsymbol{C}\boldsymbol{T} \textbf{1}$, %
$\boldsymbol{K}_{(\theta,\theta)} = \boldsymbol{M}^\top \boldsymbol{W} \boldsymbol{T}^2 \boldsymbol{M}$, %
$\boldsymbol{K}_{(\theta,\Theta)} = \boldsymbol{K}_{(\Theta,\theta)}^\top = \boldsymbol{M}^\top \boldsymbol{W} \boldsymbol{T}^2 \mathbfcal{M}$, %
$\boldsymbol{K}_{(\theta,\varphi)} = \boldsymbol{K}_{(\varphi,\theta)}^\top = \boldsymbol{M}^\top \boldsymbol{C}\boldsymbol{T} \textbf{1}$, %
$\boldsymbol{K}_{(\Theta,\Theta)} = \mathbfcal{M}^\top \boldsymbol{W} \boldsymbol{T}^2 \mathbfcal{M}$, %
$\boldsymbol{K}_{(\Theta,\varphi)} = \boldsymbol{K}_{(\varphi,\Theta)}^\top = \mathbfcal{M}^\top \boldsymbol{C}\boldsymbol{T} \textbf{1}$,
${K}_{(\varphi,\varphi)} = \mathrm{tr}(\boldsymbol{D})$,
and
$\textbf{1}$ is the $(n-m)\times1$ vector of ones.
We note that the conditional Fisher information matrix is not block diagonal, and hence the parameters are not orthogonal~\cite{Cox1987}.

Under some mild regularity conditions
the conditional maximum likelihood estimates are consistent and asymptotically normally distributed  \cite{Andersen1970}.
Thus, in large sample sizes,
\begin{align}
\label{E:normal}
\left(\begin{array}{ll}
\widehat{\beta}\\
\widehat{\boldsymbol{\phi}}\\
\widehat{\boldsymbol{\Phi}}\\
\widehat{\boldsymbol{\theta}}\\
\widehat{\boldsymbol{\Theta}} \\
\widehat{\varphi}
\end{array} \right )
\sim \mathcal{N}_{k}
\left(\begin{array}{ll}
\left(\begin{array}{ll}
\beta \\
\boldsymbol{\phi} \\
\boldsymbol{\Phi} \\
\boldsymbol{\theta} \\
\boldsymbol{\Theta} \\
\varphi
\end{array} \right ),
\boldsymbol{K}^{-1}
\end{array} \right ) ,
\end{align}
approximately, where
$\widehat{\beta}$,
$\widehat{\boldsymbol{\phi}}$,
$\widehat{\boldsymbol{\Phi}}$,
$\widehat{\boldsymbol{\theta}}$,
$\widehat{\boldsymbol{\Theta}}$ and
$\widehat{\varphi}$ are the CMLEs of
$\beta$, $\boldsymbol{\phi}$, $\boldsymbol{\Phi}$, $\boldsymbol{\theta}$, $\boldsymbol{\Theta}$ and $\varphi$, respectively.
Notice that
$\boldsymbol{K}^{-1}$
is the asymptotic covariance matrix of $\widehat{\boldsymbol{\gamma}}$.

\subsection{Confidence intervals and hypothesis testing inference}

Let $\gamma_r$ denote the $r$th component of $\boldsymbol{\gamma}$.
From \eqref{E:normal}, we have that
\begin{align*}
(\widehat{\gamma}_r - \gamma_r) \lbrace K(\widehat{\boldsymbol{\gamma}})^{rr}\rbrace^{-1/2}\, \sim\, \mathcal{N}(0,1),
\end{align*}
approximately, where $K(\widehat{\boldsymbol{\gamma}})^{rr}$ is the $r$th diagonal element of
 $\boldsymbol{K}^{-1}$. %
Let $z_{\delta}$ represent the $\delta$ standard normal quantile. A $100(1-\alpha)\%$, $0 < \alpha < 1/2$, confidence interval for $\gamma_r$, $r=1,\ldots,k$, is
\begin{align*}
\left[\widehat{\gamma}_r - z_{1-\alpha/2} \left(K(\widehat{\boldsymbol{\gamma}})^{rr}\right)^{1/2};\widehat{\gamma}_r + z_{1-\alpha/2} \left(K(\widehat{\boldsymbol{\gamma}})^{rr}\right)^{1/2}\right].
\end{align*}
Details on asymptotic confidence intervals can be found in \cite{Pawitan2001} and \cite{Davison1997}.

The test for $\mathcal{H}_0:\gamma_r=\gamma_{r}^0$ against $\mathcal{H}_1:\gamma_r \neq \gamma_{r}^0$ can be based on
the signed square root of Wald's statistic, which is given by~\cite{Pawitan2001}
\begin{align*}
z = \dfrac{\widehat{\gamma}_r-\gamma_{r}^0}{\widehat{{\rm se}}(\widehat{\gamma}_r)},
\end{align*}
where the asymptotic standard error of the $\widehat{\gamma}_r$ is
$\widehat{{\rm se}}(\widehat{\gamma}_r)= \left(K(\widehat{\boldsymbol{\gamma}})^{rr}\right)^{1/2}$.
Under $\mathcal{H}_0$, the limiting distribution of $z$ is standard normal.

It is possible to perform hypothesis testing inference using the
the likelihood ratio~\cite{Person1928},
Rao's score~\cite{Rao1948},
Wald~\cite{Wald1943},
and
gradient~\cite{Terrell2002} tests.
In large samples and under the null hypothesis, such test statistics are chi-squared distributed.

The Wald test for the presence of seasonal movements can be carried out as follows. The null and alternative
hypotheses are
\begin{align*}
\mathcal{H}_0: & (\Phi_1,\ldots,\Phi_P,\Theta_1,\ldots,\Theta_Q)^\top = \mathbf{0}_{P+Q} \quad \text{(non-seasonal),} \\
\mathcal{H}_1: & (\Phi_1,\ldots,\Phi_P,\Theta_1,\ldots,\Theta_Q)^\top \neq \mathbf{0}_{P+Q} \quad \text{(seasonal),}
\end{align*}
where $\mathbf{0}_{P+Q}$ is the $(P+Q)$-vector of zeros.
Under $\mathcal{H}_0$, there is no seasonal dynamics. Rejection of the null hypothesis indicates that seasonality must be accounted for. The Wald test statistics is
\begin{align*}
W= (\widehat{\Phi}_1,\ldots,\widehat{\Phi}_P,\widehat{\Theta}_1,\ldots,\widehat{\Theta}_Q)
\left(\boldsymbol{K}(\widehat{\boldsymbol{\gamma}})^{\Phi \Theta}\right)^{-1}
(\widehat{\Phi}_1,\ldots,\widehat{\Phi}_P,\widehat{\Theta}_1,\ldots,\widehat{\Theta}_Q)^\top,
\end{align*}
where $\boldsymbol{K}(\widehat{\boldsymbol{\gamma}})^{\Phi \Theta}$
is the $(P+Q) \times (P+Q)$ block of the inverse of Fisher's information matrix
relative to the seasonal parameters evaluated at $\hat{\boldsymbol{\gamma}}$.
Under standard regularity conditions and under the null hypothesis, $W$ is asymptotically chi-squared distribution with $P+Q$ degrees of freedom ($\chi^2_{P+Q}$).
Notice that in order to compute $W$ one only needs to estimate the non-null (seasonal) model.

\section{Model selection, diagnostic analysis and forecasting}\label{S:diagnostic}

In what follows we present some model selection criteria that can be used for model identification and present some diagnostic tools for fitted $\beta$SARMA models. Diagnostic checks can be applied to a fitted model to determine whether it fully captures the data dynamics. A fitted model that passes all diagnostic checks can then be used for out-of-sample forecasting.

\subsection{Model selection criteria}

Model selection can be based on Akaike's Information Criterion (AIC)~\cite{Akaike1974}. Since the conditional log-likelihood is additive, using the idea introduced by \cite{Pan1999} for bootstrapped likelihood cross validation, we propose the following modified AIC:
\begin{align}\label{e:maic}
{\rm MAIC} = -2 \hat{\ell}_\ast +2k,
\end{align}
where $\hat{\ell}_\ast  =  \hat{\ell} \times n/(n-m) $
and $k=p+q+P+Q+2$ is the number of parameters in the model.
When comparing models of different dimensions for different values of $m$, $\hat{\ell}_\ast$ can be interpreted as the sum of $n$ terms.
Therefore, the MAIC does not incorrectly penalize models with larger values of $m$.

The MAIC aims at estimating the expected conditional log-likelihood using a bias correction ($2k$) for the maximized conditional log-likelihood function. When $2k$ in \eqref{e:maic} is replaced by $\log(n)k$ we obtain the modified Schwarz Information Criterion (MSIC)~\cite{Schwarz1978}; when it is replaced by $\log\left[\log(n)\right]k$, the modified Hannan and Quinn Information Criterion (MHQ)~\cite{Hannan1979} is obtained. Alternative choices can be considered for the bias correcting term, such as in \cite{Sugiura1978, Hurvich1989, McQuarrie1999} or the bootstrapped versions in \cite{Shao1996,Cavanaugh1997b,Ishiguro1997,Cavanaugh1997,Shibata1997,Seghouane2010,Bayer2015-test}.

\subsection{Deviance}

The deviance is defined as twice the difference between the conditional log-likelihood evaluated at the saturated model (for which  $\tilde{\mu}_t=y_t$) and at the fitted model. That is, the deviance is given by
\begin{align*}
D=2\left[ \tilde{\ell} - \hat{\ell} \right],
\end{align*}
where $\hat{\ell}  =  \sum\limits_{t=m+1}^{n}\ell_t(\hat{\mu}_t,\hat{\varphi})$ and $\tilde{\ell}  =  \sum\limits_{t=m+1}^{n}\ell_t(y_t,\hat{\varphi})$.
When the fitted model is correctly specified, $D$ is approximately distributed as $\chi^2_{n-m-k}$~\cite{Benjamin2003, Kedem2002}. It is customary to divide the deviance by $n-m-k$. There is evidence of incorrect model specification when $D/(n-m-k)$ is considerably larger than one~\cite{Myers2010}.

\subsection{Residuals}

Residual analysis is important for determining whether the model at hand provides a good fit~\cite{Kedem2002}. Visual inspection of a time series residuals plot is an indispensable first step when assessing goodness-of-fit~\cite{Box2008}. Various types of residuals are currently available~\cite{Mauricio2008}. Since the model we propose is an extension of the beta regression model~\cite{Ferrari2004} for time series analysis, the residual used in beta regression diagnostics can also be used here. For details on residuals and diagnostics tools in beta regression models, see \cite{espinheira2008,espinheira2008b}. For details on time series model residuals, see~\cite{Kedem2002}.

At the outset, we define the following standardized residual:
\begin{align*}
\hat{r}_t^1 = \frac{y_t-\widehat{\mu}_t}{\sqrt{\widehat{\rm Var}(y_t)}} =
\frac{y_t - \widehat{\mu}_t}{\sqrt{V(\widehat{\mu}_t)/(1+\widehat{\varphi})}}.
\end{align*}
Considering the predictor scale, we define the following standardized residual 2:
\begin{align*}
\hat{r}_t^2 =
\frac{g(y_t)-\widehat{\eta}_t}{\sqrt{ \left(g'(\widehat{\mu}_t) \right)^2 V(\widehat{\mu}_t)/(1+\widehat{\varphi}) }} .
\end{align*}
Using a Taylor series expansion, in \cite{Rocha2009} is shown that ${\rm Var}\left(g(y_t) \right) \approx \left(g'({\mu}_t) \right)^2 V({\mu}_t)/(1+{\varphi})$.

A more sophisticated residual is the standardized weighted residual introduced by \cite{espinheira2008b}, which is given by
\begin{align*}
\hat{r}_t^w = \frac{y_t^\ast-\widehat{\mu}_t^\ast}{\sqrt{\widehat{\rm Var}(y_t^\ast)}} =
\frac{y_t^\ast-\widehat{\mu}_t^\ast}{\sqrt{ \psi'(\widehat{\mu}_t \widehat{\varphi}) + \psi'\left[(1-\widehat{\mu}_t) \widehat{\varphi} \right] }}.
\end{align*}
The authors have shown that ${\rm Var}(y_t^\ast)= \psi'({\mu}_t {\varphi}) + \psi'\left[(1-{\mu}_t) {\varphi} \right]$.
Under correct model specification,
the distribution of such a residual is approximately standard normal.

\subsection{White noise tests}\label{s:noise-test}

When the model is correctly specified the residuals are expected to behave as white noise, i.e., they are expected to be serially uncorrelated and follow a zero mean and constant variance process~\cite{Kedem2002}. Let $\hat{r}_{m+1}^w, \ldots, \hat{r}_n^w$ be the standardized weighted residuals obtained from the fitted model. The residual autocorrelation function (ACF) is
\begin{align*}
\widehat{\rho}(i)= \frac{\sum_{t=m+1}^{n-i}(\hat{r}_t^w-\bar{r}^w)(\hat{r}_{t+i}^w-\bar{r}^w)}{\sum_{t=m+1}^{n-i}(\hat{r}_t^w-\bar{r}^w)^2}, \quad i=0,1,\ldots,
\end{align*}
where $\bar{r}^w = (n-m)^{-1} \sum_{t=m+1}^{n}\hat{r}_t^w$.
When $i>1$ and $n$ is sufficiently large, the distribution of $\widehat{\rho}(i)$ is approximately normal with zero mean and variance $1/(n-m)$~\cite{Kedem2002, Anderson1942, Box2008}. Hence, plots the residuals ACF with horizontal lines at $\pm 1.96/\sqrt{(n-m)}$ can be used for assessing whether the residuals display white noise behavior~\cite{Kedem2002}. It is expected that $95\%$ of residuals autocorrelations lie inside the interval $[- 1.96/\sqrt{(n-m)}, 1.96/\sqrt{(n-m)}]$.

It is also possible to test the null hypothesis that the first $b$ residual autocorrelations are equal to zero. To that end, the following test statistic can be used~\cite{Ljung1978}:
\begin{align*}
Q_1=(n-m)(n-m+2)\sum_{i=1}^b \frac{\left[\widehat{\rho}(i)\right]^2}{n-m-i}.
\end{align*}
Alternatively, it is possible to base the test statistic on residual partial autocorrelations~\cite{Monti1994}:
\begin{align*}
Q_2 = (n-m)(n-m+2)\sum_{i=1}^b \frac{\left[p(i)\right]^2}{n-m-i},
\end{align*}
where $p(i)$ is the $i$th partial residual autocorrelation.

The critical value used in either test is obtained from the test statistic asymptotic null distribution, namely $\chi^2_{b-p-q-P-Q}$. The null hypothesis is rejected at nominal level $\alpha$ if $Q_j > \chi^2_{1-\alpha, b-p-q-P-Q}$, $j=1,2$, where $Q_j > \chi^2_{1-\alpha, b-p-q-P-Q}$ is the $1-\alpha$ $\chi^2_{b-p-q-P-Q}$ quantile.
Based on pilot simulations and on a rule-of-thumb available in the literature~\cite{Hyndman2013},
we suggest using
$b=\max(10,2S)$,
where $S$ is the seasonality frequency.

\subsection{Forecasting}

Estimates of $\mu_t$, $\widehat{\mu}_t$, for $t=m,\ldots,n$ (in sample), can be obtained by replacing $\gamma$ by its CMLE, $\widehat{\gamma}$, and $r_t$ by $g(y_t)-g(\widehat{\mu}_t)$ in the Equation~\eqref{E:bsarma}. The $h$ step ahead forecast, $h=1,2,\ldots$, can be computed as
\begin{align*}%
\widehat{\mu}_{n+h}=g^{-1} \left(
 \widehat{\beta}
+
\sum\limits_{i=1}^{p}\widehat{\phi}_i  \left[ g(y_{n+h-i}) \right]
-
\sum\limits_{j=1}^{q}\widehat{\theta}_j \left[ r_{n+h-j} \right]
+
\sum\limits_{I=1}^{P}\widehat{\Phi}_I  \left[ g(y_{n+h-IS}) \right]
-
\sum\limits_{J=1}^{Q}\widehat{\Theta}_J \left[ r_{n+h-JS} \right] \right. \nonumber\\
\left.
 -
\sum\limits_{i=1}^{p}\sum\limits_{I=1}^{P}\widehat{\phi}_i\widehat{\Phi}_I
\left[ g(y_{n+h-(i+IS)}) \right]
+
\sum\limits_{j=1}^{q}\sum\limits_{J=1}^{Q}\widehat{\theta}_j\widehat{\Theta}_J
\left[ r_{n+h-(j+JS)} \right]
\right),
\end{align*}
where
\begin{align*}
\left[ g(y_{t}) \right]  =\left\{\begin{array}{rc}
g(\widehat{\mu}_{t}),&\textrm{if}\;\;\; t>n,\\
g(y_{t}), &\textrm{if}\;\;\; t\leq n,
\end{array}\right.
\quad \text{and} \quad
\left[ r_{t} \right] =\left\{\begin{array}{rc}
0,&\textrm{if}\;\;\; t>n,\\
g(y_t) - g(\widehat{\mu}_t), &\textrm{if}\;\;\; t\leq n.
\end{array}\right.
\end{align*}

\section{Numerical evaluation}\label{S:simulation}

We performed Monte Carlo simulations to evaluate the finite sample performance of the CMLE and also the accuracy of white noise testing inference. We used $R=10,000$ Monte Carlo replications and the following sample sizes: $n=50, 100, 200, 500$. In each Monte Carlo replication we generate a vector of $n$ occurrences of the variable $y_t$ from the $\beta$SARMA model given in \eqref{E:bsarma} with logit link. The parameter values are presented in Table~\ref{t:simu} along with the numerical results. We report the mean of all estimates, and also estimates of the bias, estimated relative bias (RB), standard deviation (SD), and mean square error (MSE). All conditional log-likelihood maximizations were carried out using the BFGS quasi-Newton method with analytical first derivatives. Starting values for the autoregressive parameters were obtained by regressing $g(y_t)$ on $g(y_{t-1}), \ldots, g(y_{t-p}), g(y_{t-(p+1)}),\ldots,g(y_{t-(p+P)})$, and the all moving average parameters were set equal to zero at the beginning of the conditional log-likelihood maximizations. The simulations were performed using the \textsc{R} statistical computing environment~\cite{R2012}.

\begin{table}[p]
\begin{center}
\caption{Simulation results on point estimation, $\beta$SARMA$(1,1)\times(1,1)_{12}$} \label{t:simu}
\begin{tabular}{lrrrrrr}
\hline
 	&	$\beta$	&	$\phi_1$	&	$\Phi_1$	&	$\theta_1$	&	$\Theta_1$	&	$\varphi$	\\
Parameters	& $	-1.000	$ & $	-0.500	$ & $	0.300	$ & $	0.400	$ & $	-0.350	$ & $	120.000	$ \\
\hline
\multicolumn{7}{c}{$n=50$}\\
\hline
Mean	& $	-0.7494	$ & $	-0.4725	$ & $	0.4626	$ & $	0.4415	$ & $	-0.1233	$ & $	99.6751	$ \\
Bias	& $	-0.2506	$ & $	-0.0275	$ & $	-0.1626	$ & $	-0.0415	$ & $	-0.2267	$ & $	20.3249	$ \\
RB (\%)	& $	25.0592	$ & $	5.5057	$ & $	-54.1886	$ & $	-10.3711	$ & $	64.7802	$ & $	16.9374	$ \\
SD	& $	0.3296	$ & $	0.2065	$ & $	0.2189	$ & $	0.2542	$ & $	0.3357	$ & $	29.6409	$ \\
MSE	& $	0.1714	$ & $	0.0434	$ & $	0.0744	$ & $	0.0664	$ & $	0.1641	$ & $	1291.6873	$ \\
\hline
\multicolumn{7}{c}{$n=100$}\\
\hline
Mean	& $	-0.8698	$ & $	-0.4807	$ & $	0.3816	$ & $	0.4181	$ & $	-0.2464	$ & $	96.9399	$ \\
Bias	& $	-0.1302	$ & $	-0.0193	$ & $	-0.0816	$ & $	-0.0181	$ & $	-0.1036	$ & $	23.0601	$ \\
RB (\%)	& $	13.0189	$ & $	3.8569	$ & $	-27.1955	$ & $	-4.5257	$ & $	29.6053	$ & $	19.2167	$ \\
SD	& $	0.2425	$ & $	0.1343	$ & $	0.1604	$ & $	0.1487	$ & $	0.1940	$ & $	18.2091	$ \\
MSE	& $	0.0758	$ & $	0.0184	$ & $	0.0324	$ & $	0.0224	$ & $	0.0484	$ & $	863.3406	$ \\
\hline
\multicolumn{7}{c}{$n=200$}\\
\hline
Mean	& $	-0.9363	$ & $	-0.4855	$ & $	0.3371	$ & $	0.4132	$ & $	-0.3018	$ & $	105.2458	$ \\
Bias	& $	-0.0637	$ & $	-0.0145	$ & $	-0.0371	$ & $	-0.0132	$ & $	-0.0482	$ & $	14.7542	$ \\
RB (\%)	& $	6.3663	$ & $	2.8974	$ & $	-12.3528	$ & $	-3.3068	$ & $	13.7660	$ & $	12.2952	$ \\
SD	& $	0.1830	$ & $	0.0928	$ & $	0.1238	$ & $	0.0994	$ & $	0.1360	$ & $	13.7747	$ \\
MSE	& $	0.0375	$ & $	0.0088	$ & $	0.0167	$ & $	0.0101	$ & $	0.0208	$ & $	407.4279	$ \\
\hline
\multicolumn{7}{c}{$n=500$}\\
\hline
Mean	& $	-0.9842	$ & $	-0.4947	$ & $	0.308	$ & $	0.404	$ & $	-0.3385	$ & $	111.9313	$ \\
Bias	& $	-0.0158	$ & $	-0.0053	$ & $	-0.008	$ & $	-0.004	$ & $	-0.0115	$ & $	8.0687	$ \\
RB (\%)	& $	1.5808	$ & $	1.0617	$ & $	-2.6824	$ & $	-0.9898	$ & $	3.2979	$ & $	6.7240	$ \\
SD	& $	0.1205	$ & $	0.0563	$ & $	0.0805	$ & $	0.0606	$ & $	0.0825	$ & $	9.4855	$ \\
MSE	& $	0.0148	$ & $	0.0032	$ & $	0.0066	$ & $	0.0037	$ & $	0.0069	$ & $	155.0802	$ \\
\hline
\end{tabular}
\end{center}%
\caption{Null rejection rates of the Ljung-Box and Monti tests} \label{t:simu-ljung}
\begin{center}
\begin{tabular}{lrrrr}
\hline
	& $	n=50	$ & $	n=100	$ & $	n=200	$ & $	n=500	$ \\
\hline
\multicolumn{5}{c}{$\alpha=10\%$}\\
\hline
Ljung-Box	& $ 	10.77	$ & $	12.56	$ & $	11.27	$ & $	11.97	$ \\
Monti	& $	9.55	$ & $	13.59	$ & $	12.53	$ & $	12.49	$ \\
\hline
\multicolumn{5}{c}{$\alpha=5\%$}\\
\hline
Ljung-Box	& $	6.29	$ & $	7.24	$ & $	6.10	$ & $	6.15	$ \\
Monti	& $ 	3.94	$ & $	6.98	$ & $	6.63	$ & $	6.39	$ \\
\hline
\multicolumn{5}{c}{$\alpha=1\%$}\\
\hline
Ljung-Box	& $	1.97	$ & $	2.36	$ & $	1.58	$ & $	1.52	$ \\
Monti	& $ 	0.53	$ & $	1.26	$ & $	1.51	$ & $	1.46	$ \\
\hline
\end{tabular}
\end{center}
\end{table}

The results in Table~\ref{t:simu} show that the estimator of the autoregressive parameter $\phi_1$ is nearly unbiased. Similar to what happens in beta regressions~\cite{simas2010}, the precision parameter estimator displays some small sample bias. The remaining estimators display substantial bias when the sample size is small. Such biases become smaller as the sample size grows.

\begin{figure}[t]
\begin{center}
\subfigure[Scenario 1]
{\label{f:poder-ma}\includegraphics[width=0.47\textwidth]{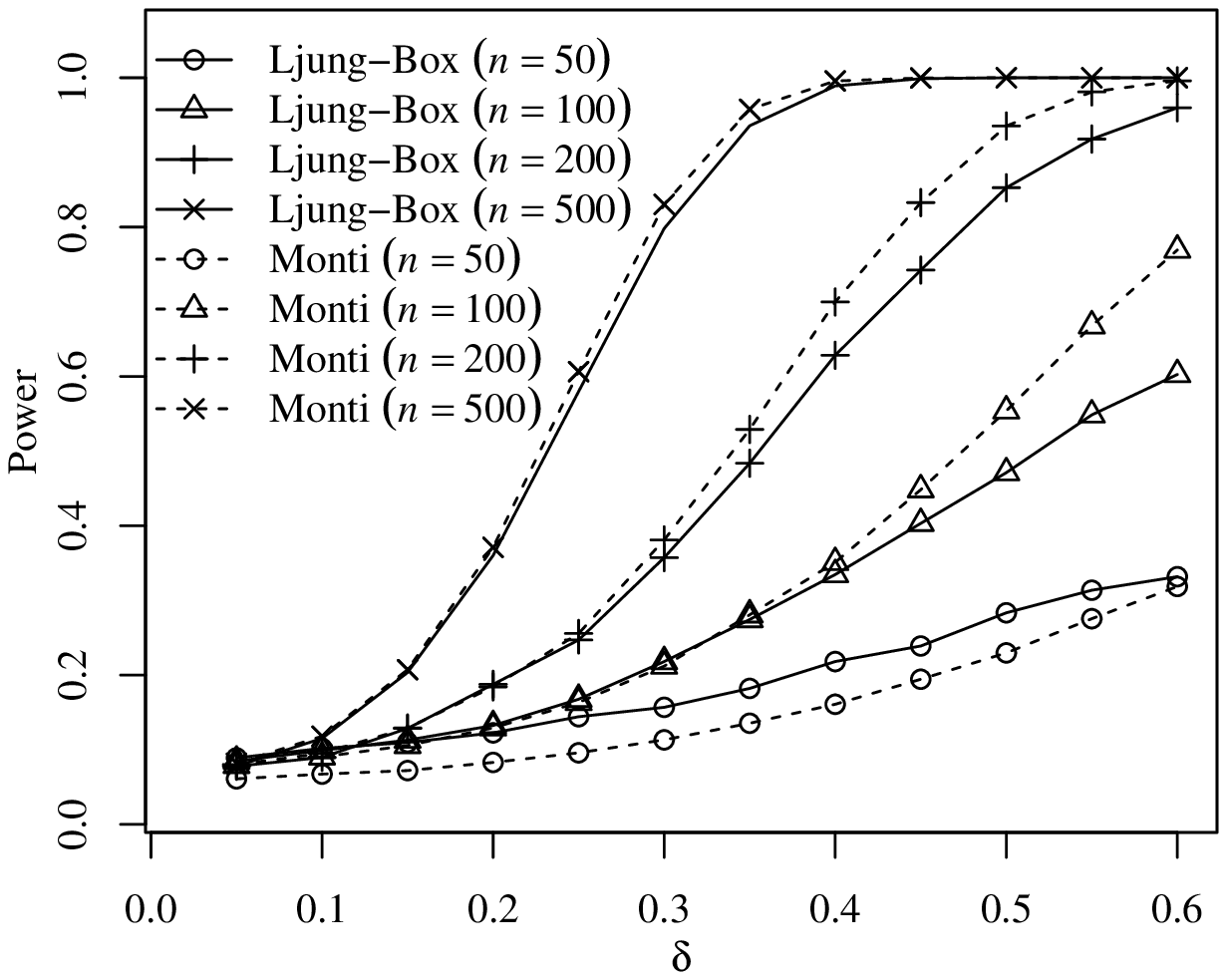}}
\subfigure[Scenario 2]
{\label{f:poder-ar}\includegraphics[width=0.47\textwidth] {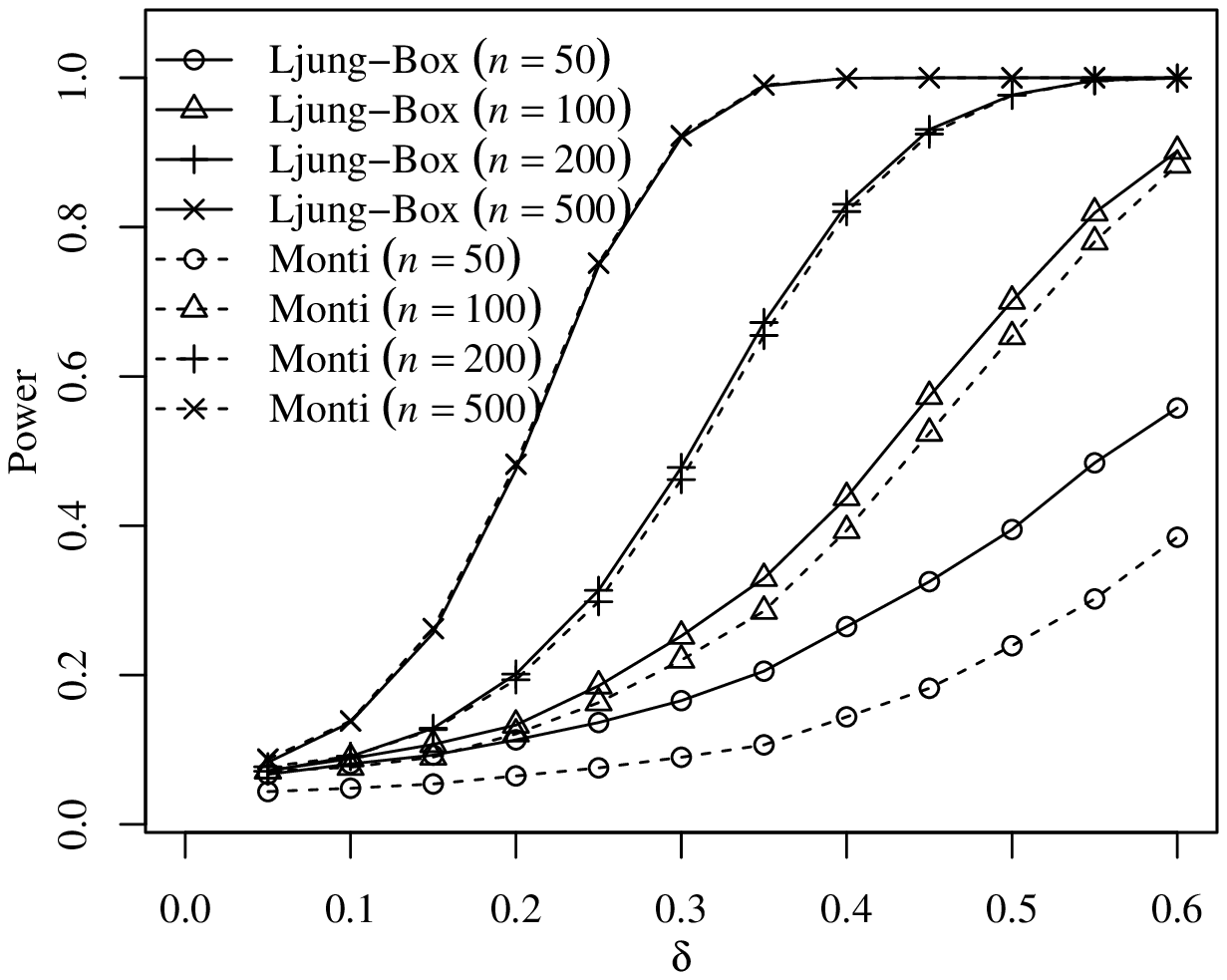}}
\caption{
Nonnull rejection rates (powers) of Ljung-Box and Monti tests; $\alpha=5\%$.
}\label{f:poder}
\end{center}
\end{figure}

We also estimated the null rejection rates (sizes) of the Ljung-Box and Monti-type tests presented in Section~\ref{s:noise-test}, based on the standardized weighted residual ($\hat{r}_t^w$). The simulation setup was the same as in the previous set of simulations. The tests nominal levels are $10\%$, $5\%$ and $1\%$ and we used $b=\max(10,2S)$, with $S=12$. The results are displayed in Table~\ref{t:simu-ljung}. They show that the Monti test typically outperforms the Ljung-Box test.
We have also used Monte Carlo simulation to evaluate the tests nonnull rejection rates (powers).
Figure~\ref{f:poder} presents such rates
for two scenarios:
\begin{description}
\item[Scenario 1:] Data generation was carried out from $\beta$SARMA(1,1)(1,1) with parameters equal to
$\beta=-1.00$,
$\phi_1=-0.50$,
$\Phi_1=0.30$,
$\theta_1=\delta$,
$\Theta_1=-0.35$,
and
$\varphi=120$,
and fitted model was $\beta$SARMA(1,0)(1,1).

\item[Scenario 2:] Data generation was carried out from $\beta$SARMA(2,1)(1,1) with parameters equal to
$\beta=-1.00$,
$\phi_1=-0.50$,
$\phi_2=-\delta$,
$\Phi_1=0.30$,
$\theta_1=0.40$,
$\Theta_1=-0.35$,
and
$\varphi=120$,
and fitted model was $\beta$SARMA(1,1)(1,1).
\end{description}
In both scenarios, $\delta = 0.05,0.10,\ldots,0.55,0.60$.
As expected, the tests become more powerful
as $\delta$ moves away from zero
and as the sample size increases.
In small samples the Ljung-Box test is more powerful than the Monti test; in
large samples, however, they are nearly equally powerful.
Overall, both tests seem to work well.

\FloatBarrier

\section{Empirical application}\label{S:application}

\begin{figure}[p]
\begin{center}
\subfigure[RH series]
{\includegraphics[width=0.47\textwidth]{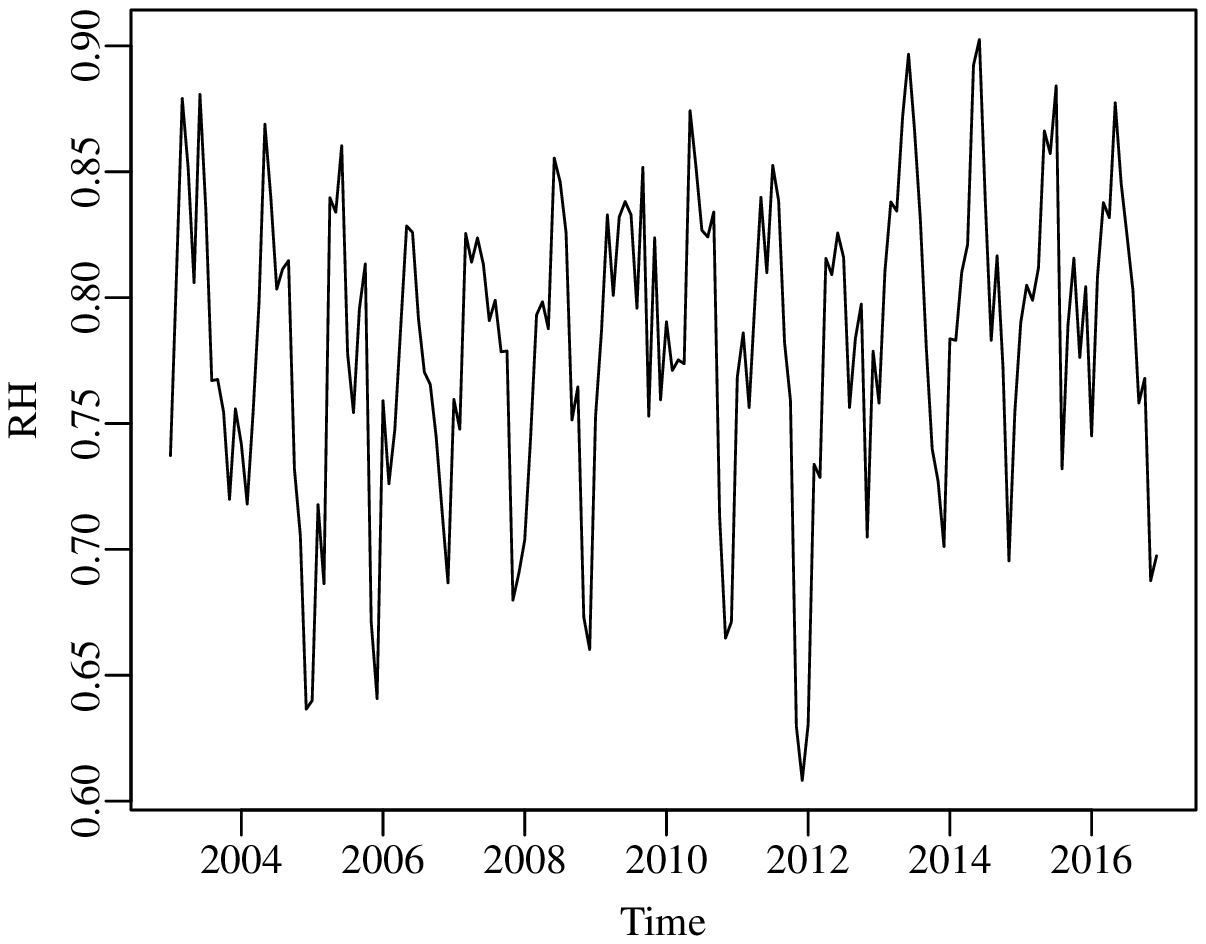}}
\subfigure[Seasonality]
{\includegraphics[width=0.47\textwidth] {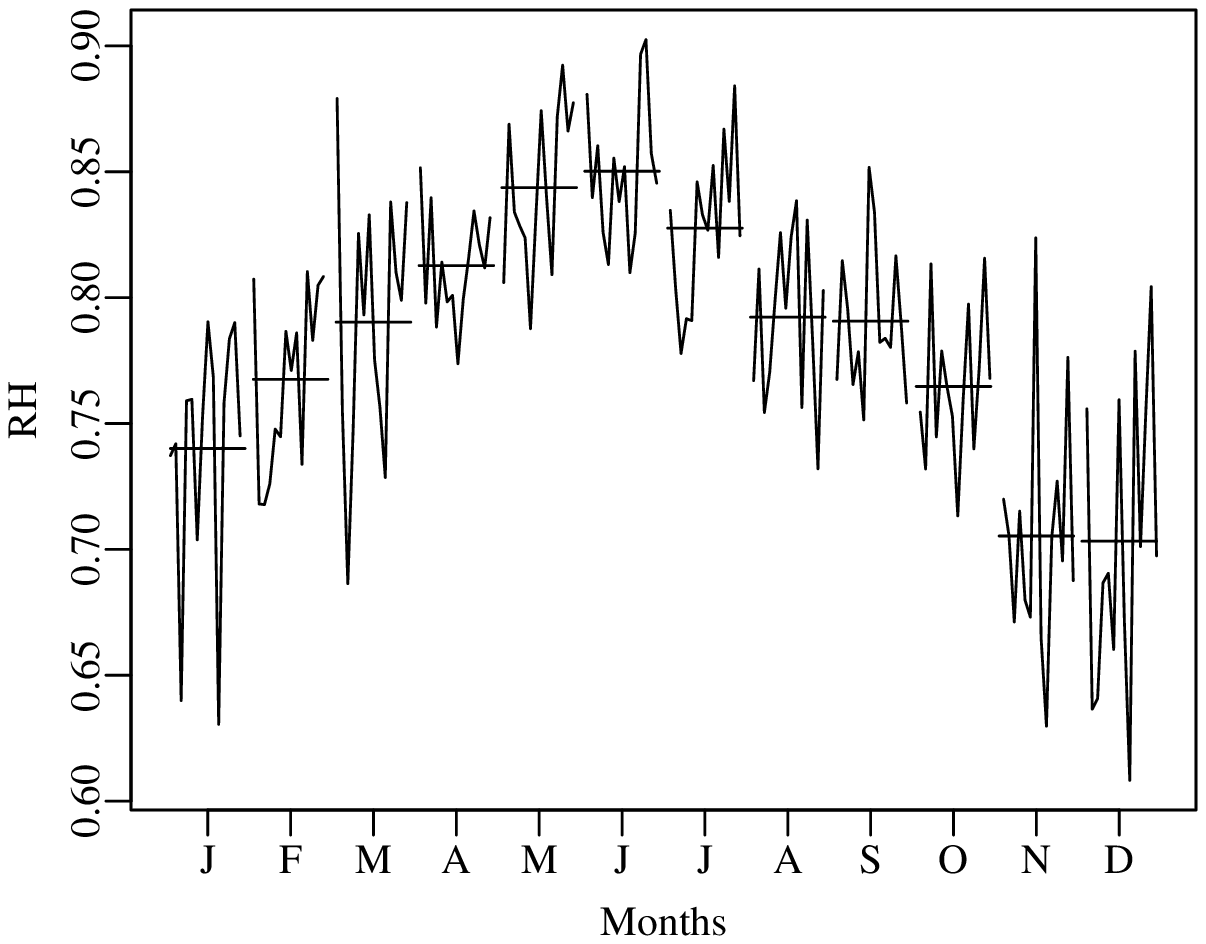}}
\subfigure[ACF]
{\includegraphics[width=0.47\textwidth]{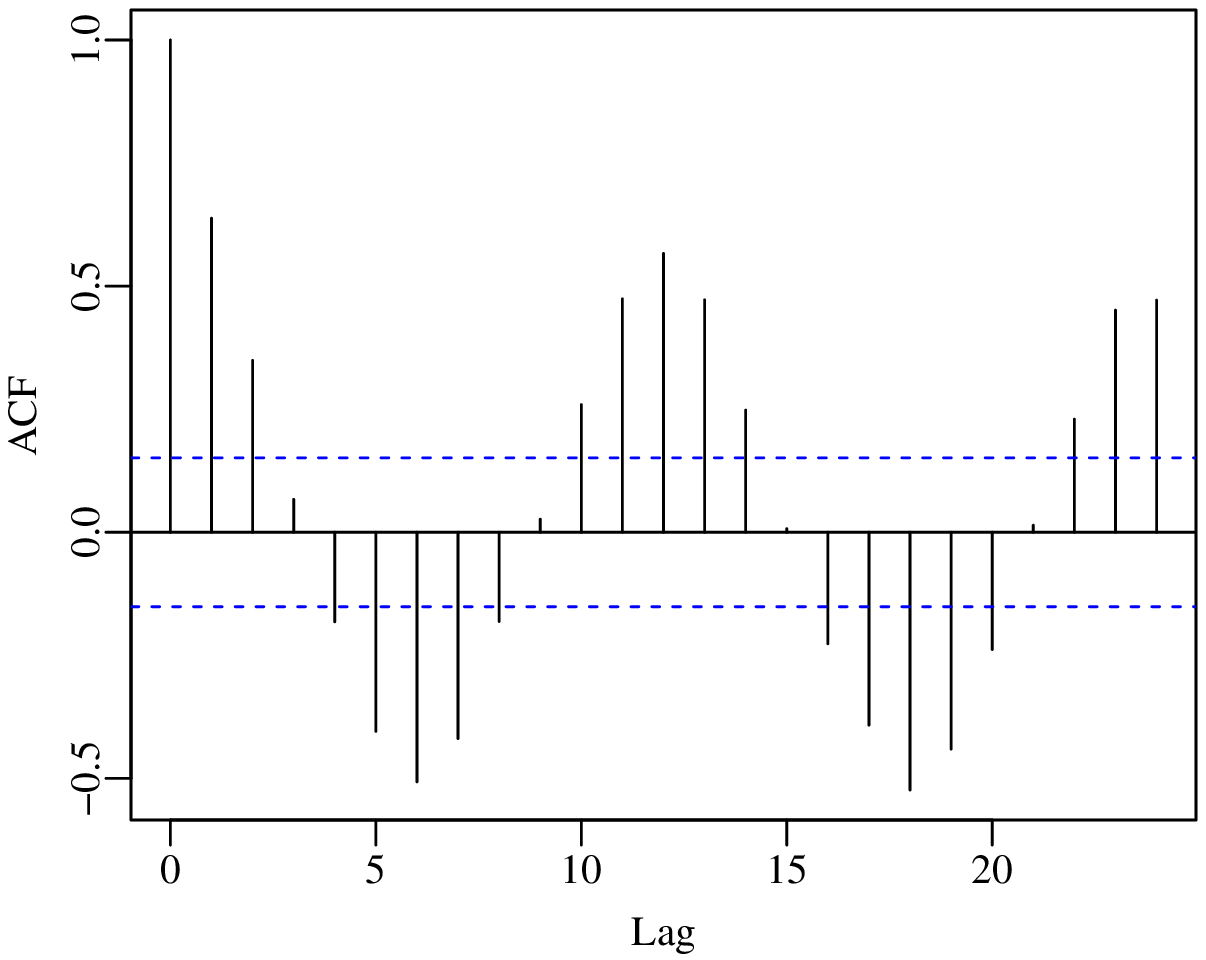}}
\subfigure[PACF]
{\includegraphics[width=0.47\textwidth] {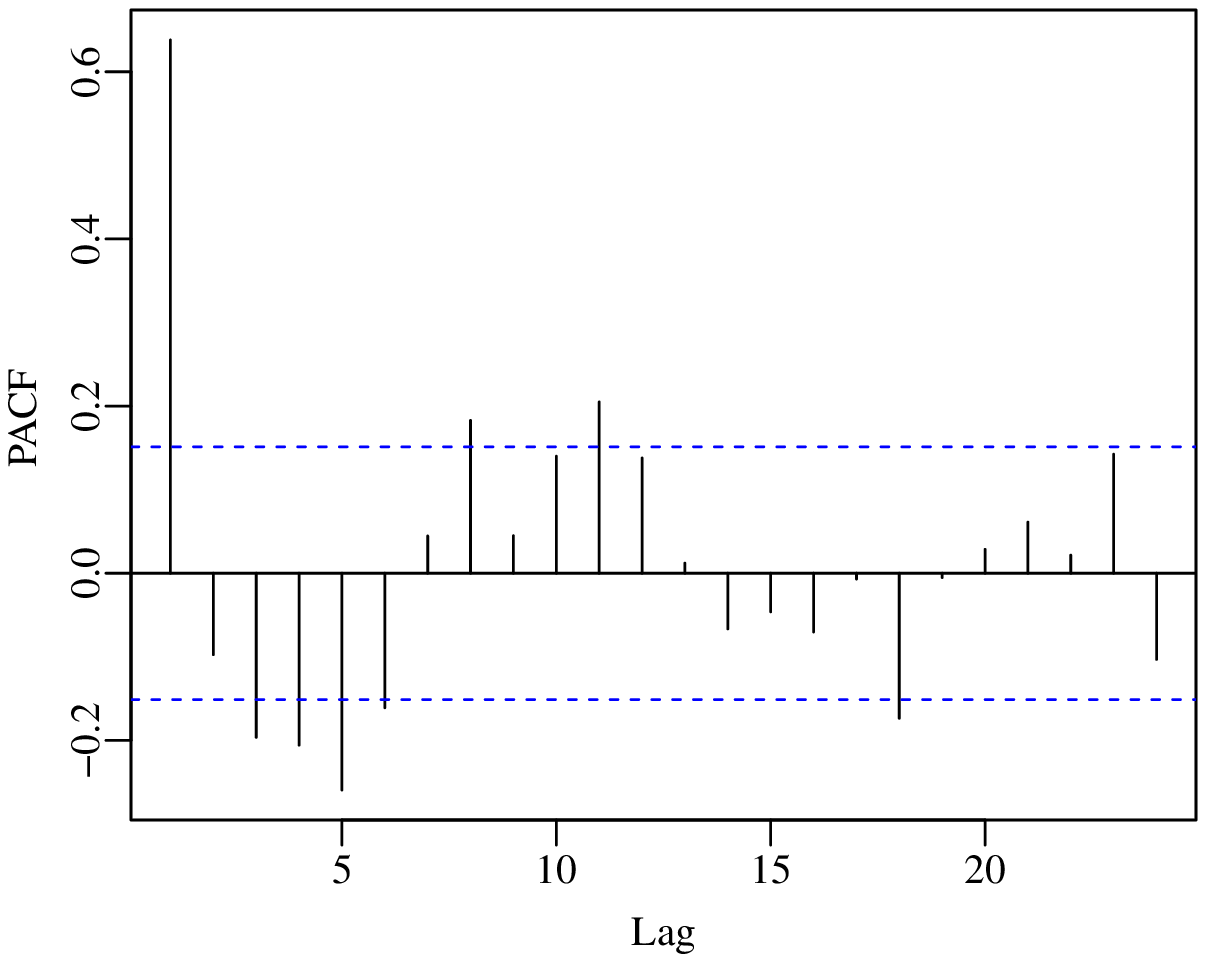}}
\caption{
Time series of relative humidity of air in Santa Maria, Brazil.
}\label{f:ur}
\end{center}
\end{figure}

We shall now model data on air relative humidity (RH) in Santa Maria, RS, Brazil. The data consist of monthly averages from January 2003 through
October 2017 %
and were obtained from Banco de Dados Meteorol\'ogicos para Ensino e Pesquisa do Instituto Nacional de Meteorologia (INMET) \cite{inmet}.
The last ten observations were removed from the data and used for forecasting evaluation.
Figure~\ref{f:ur} contains the time series data plot and also plots of the sample autocorrelation (ACF) and sample partial autocorrelation (PACF) functions. The sinusoidal patterns in both correlograms are indicative of seasonal dynamics.

The importance of modeling and forecasting environmental variables such as RH is largely discussed in literature~\cite{Tamerius2013,Dietz1976,Dowel2001,Grassly2006}. In particular, humidity seasonality has been linked to several infectious diseases~\cite{Tamerius2013}. It is thus important that seasonal fluctuations be properly modeled and that accurate forecasts are available to policymakers.

We selected the $\beta$SARMA$(1,0)\times(1,1)_{12}$ model for the data at hand. Parameter estimates, standard errors, $z$ statistics and their $p$-values are given in Table~\ref{t:fitted}. Some diagnostic measures and additional test statistics are also included in the table. Notice that the seasonality test $p$-value is quite small, which is suggestive of strong seasonal dynamics. In addition, the Ljung-Box and Monti tests
based on the standardized weighted residual
do not reject the null hypothesis that the first $b$ residual autocorrelations are equal to zero. Hence, the model seems to be correctly specified.

\begin{table}[t]
\caption{Fitted $\beta$SARMA$(1,0)\times(1,1)_{12}$ model} \label{t:fitted}
\begin{center}
\begin{tabular}{lcccc}
\hline
& estimate & std.\ error &  z stat. & $p\text{-value}$ \\
\hline
$\beta$   &   $0.1057$ &   $ 0.0365$ & $ 2.8973$ & $0.0038$\\
$\phi_1$   &   $0.3834$ &   $ 0.0437$ & $8.7657 $ & $<0.0001$\\
$\Phi_1$    &   $0.8615$ &   $ 0.0586$ & $14.7071$ & $<0.0001$\\
$\Theta_1$ &   $0.5668$ &   $ 0.0709$ & $7.9892 $ & $<0.0001$\\
$\varphi$  &   $98.3114$&   $ 11.1297$& $ 8.8332$ & $<0.0001$\\
\hline
\multicolumn{5}{c}{Log-likelihood $= 298.9695$ }\\
\multicolumn{5}{c}{Deviance $= 153.5969$}\\
\multicolumn{5}{c}{MAIC $= -585.9390$ \quad \quad MSIC = $-567.1952$}\\
\multicolumn{5}{c}{Seasonality test: $W = 265.2603$ ($p\text{-value} < 0.0001$) }\\
\multicolumn{5}{c}{Ljung-Box test: $Q_1 = 23.555$ ($p\text{-value} = 0.2624$) }\\
\multicolumn{5}{c}{Monti test: $Q_2 = 22.728$ ($p\text{-value} = 0.3023$) }\\
\hline
\end{tabular}
\end{center}%
\end{table}

\begin{figure}[p]
\begin{center}
\subfigure[Observed versus fitted]
{\label{f:obs_fit}\includegraphics[width=0.47\textwidth]{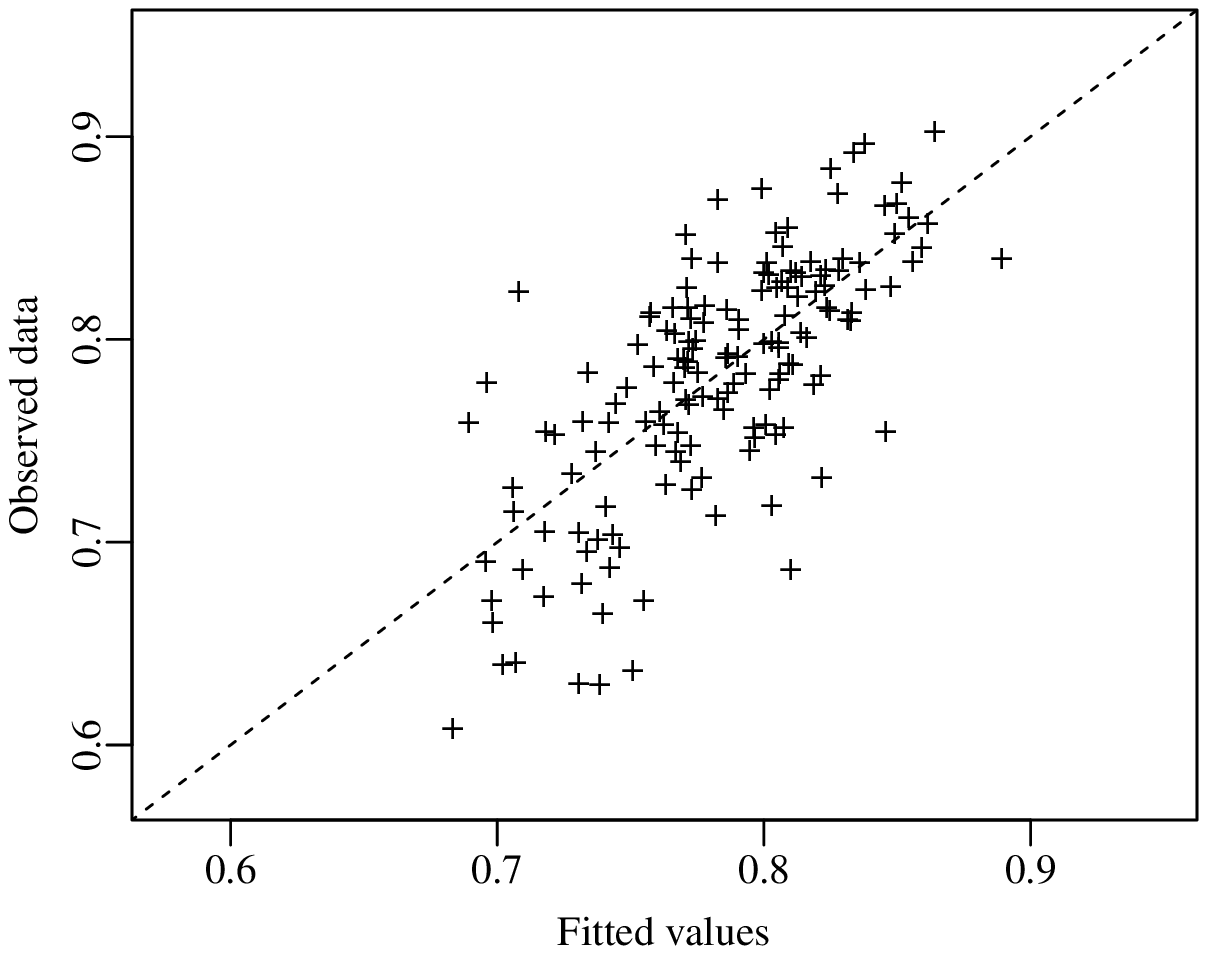}}
\subfigure[Standardized weighted residual]
{\label{f:resid_ind}\includegraphics[width=0.47\textwidth]{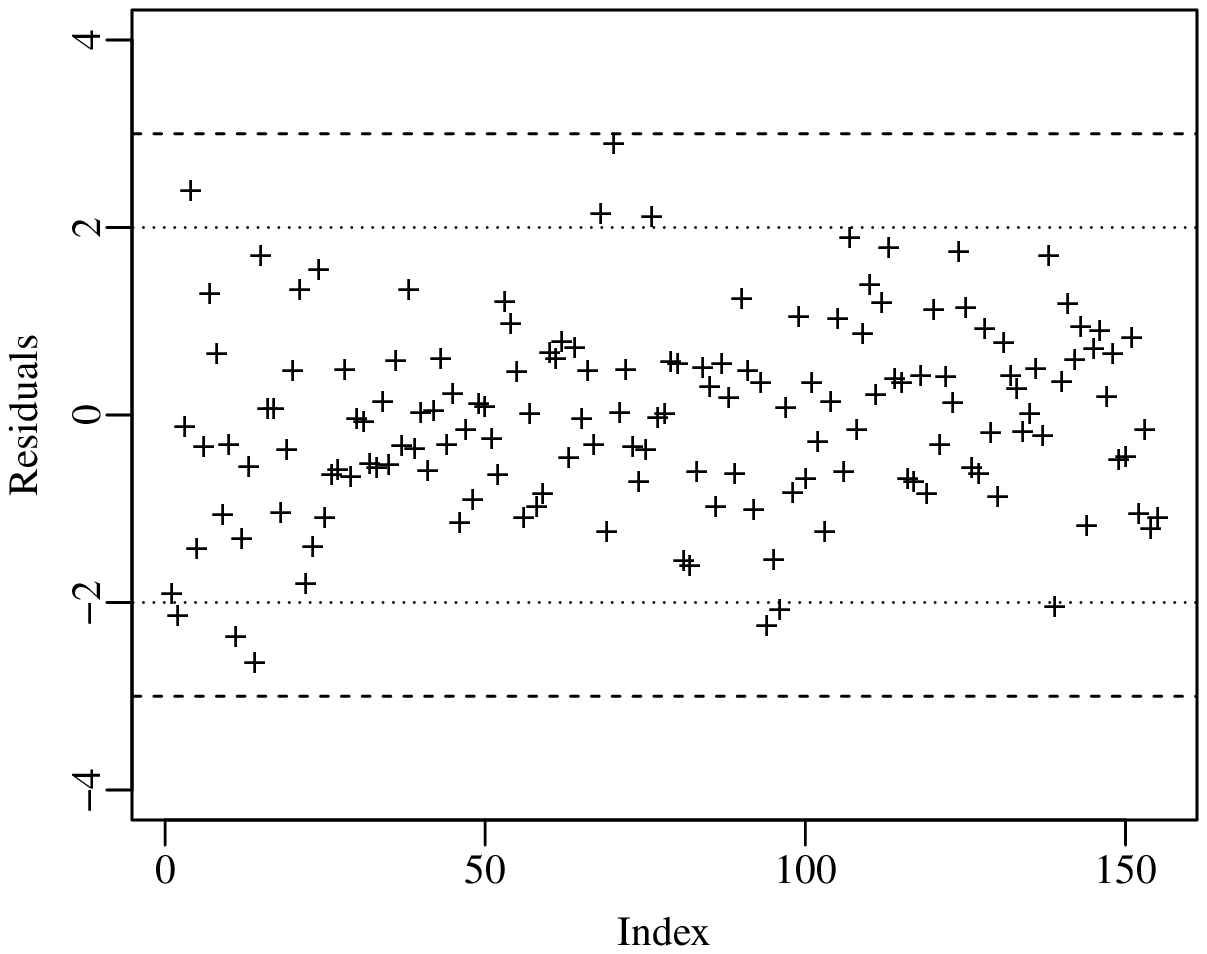}}
\subfigure[Residual ACF]
{\label{f:fac_resid}\includegraphics[width=0.47\textwidth]{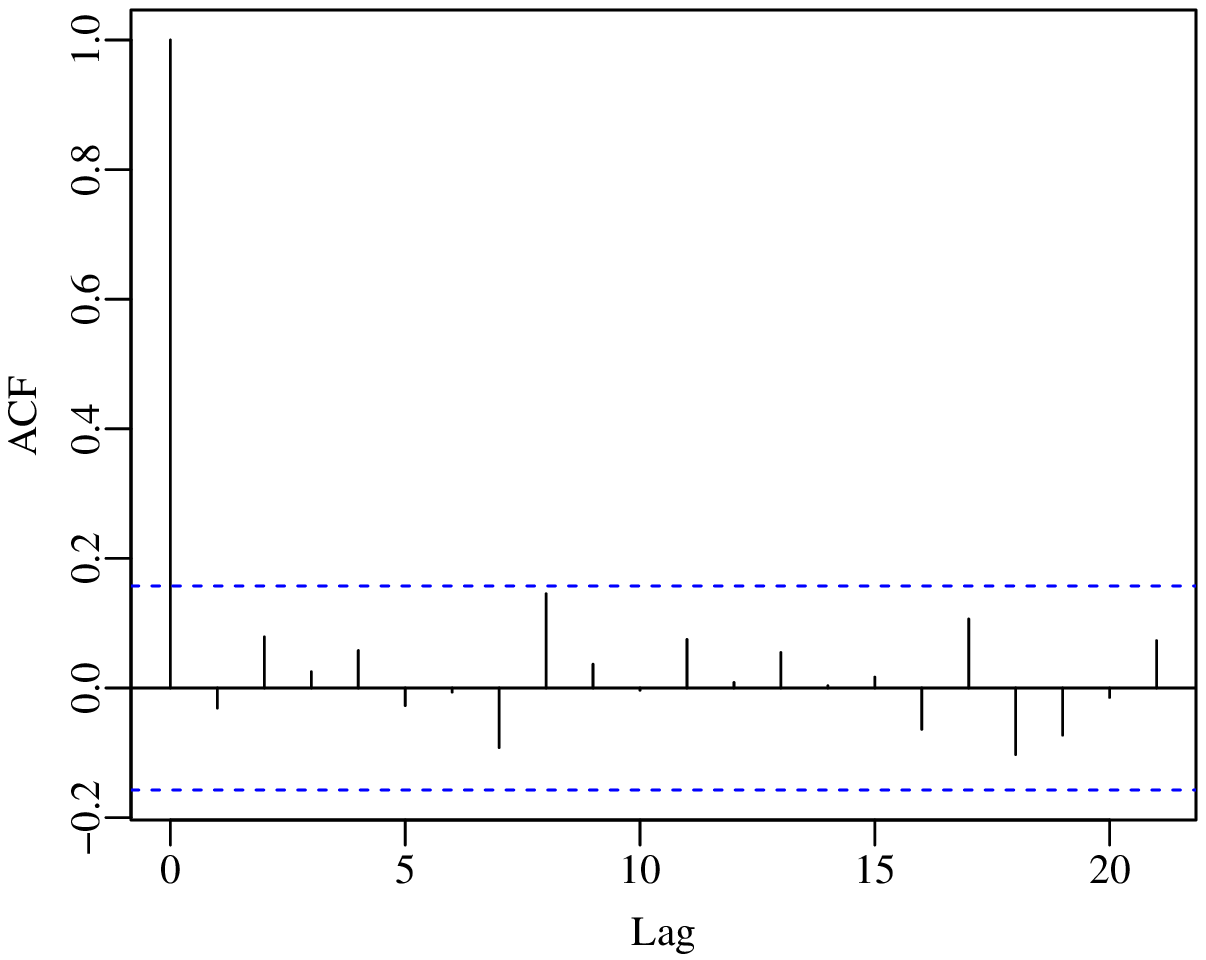}}
\subfigure[Residual PACF]
{\label{f:facp_resid}\includegraphics[width=0.47\textwidth] {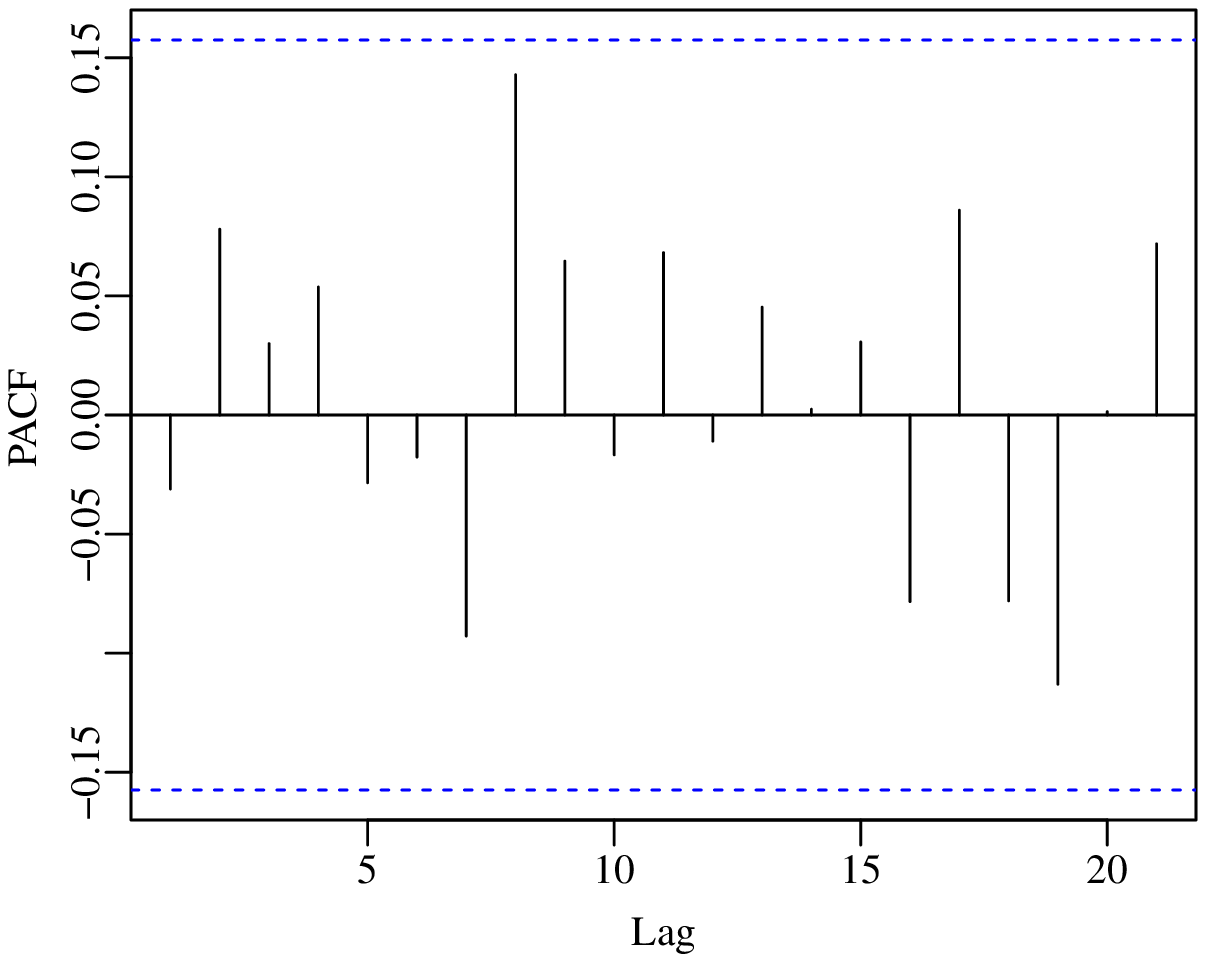}}
\subfigure[QQ-plot]
{\label{f:qqplot}\includegraphics[width=0.47\textwidth] {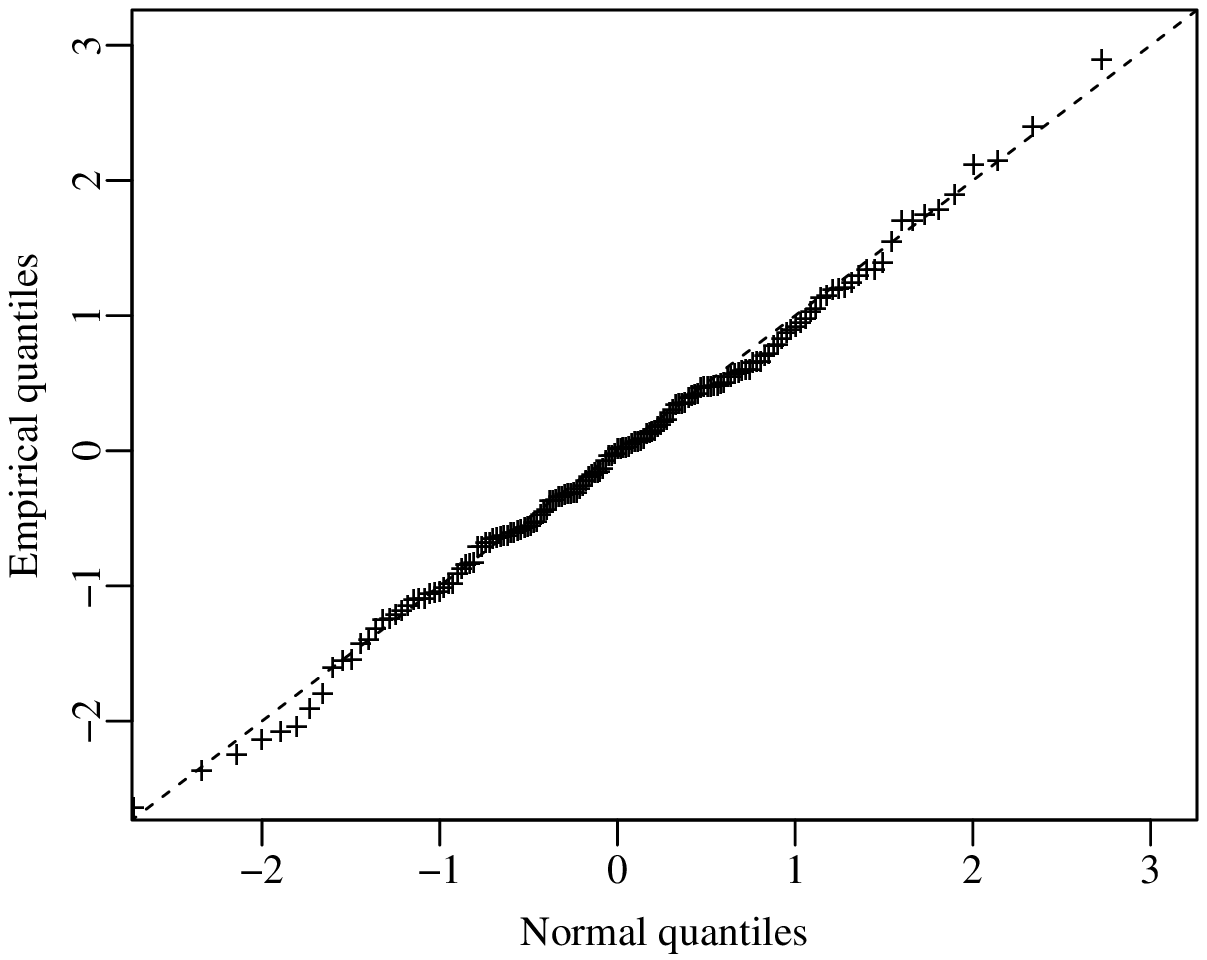}}
\subfigure[Residual distribution]
{\label{f:densitys_resid}\includegraphics[width=0.47\textwidth]{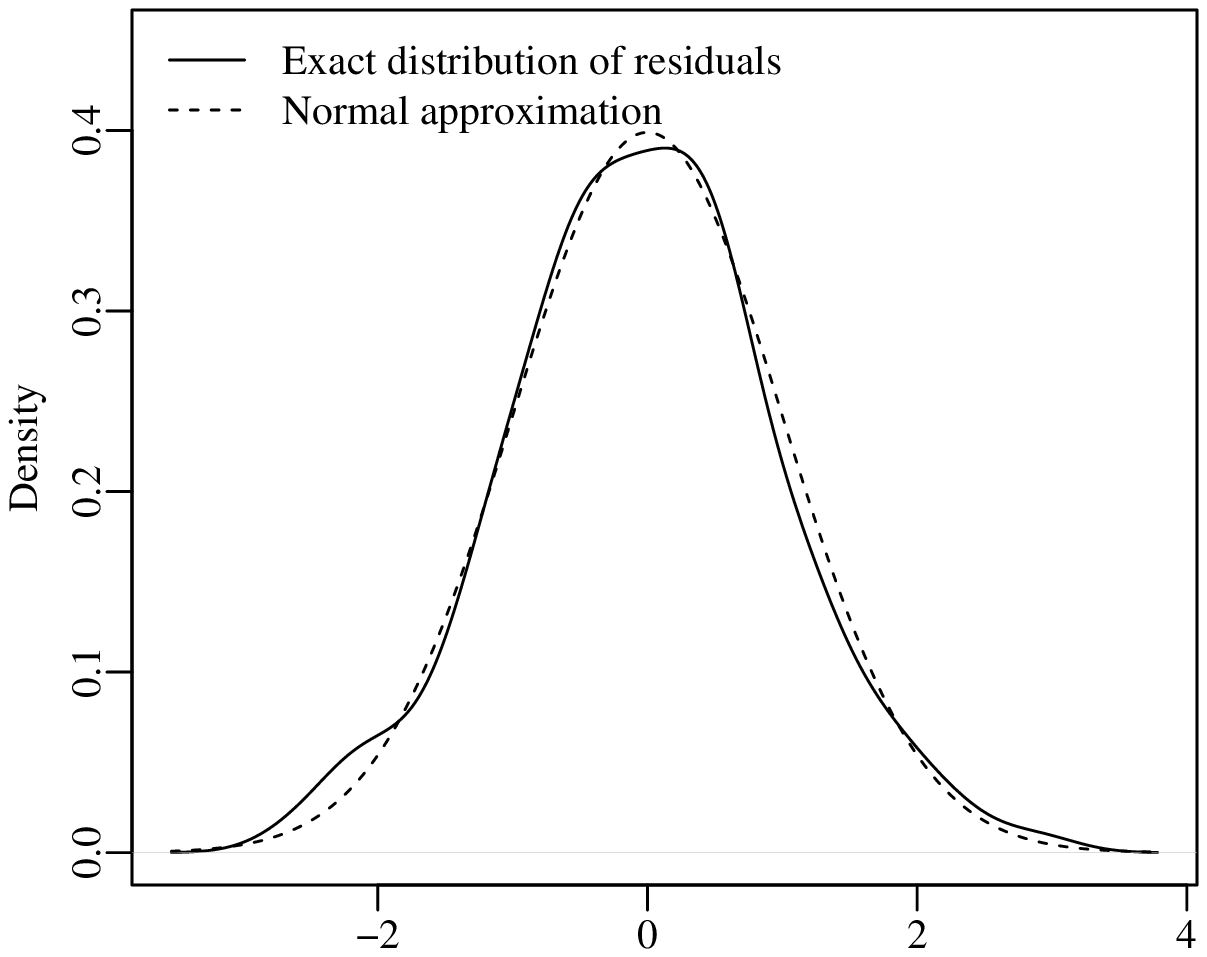}}
\caption{
Diagnostic plots based on standardized weighted residuals ($r_t^w$).
}\label{f:diag}
\end{center}
\end{figure}

\begin{figure}[t]
\begin{center}
\subfigure[Fitted $\beta$SARMA(1,0)$\times$(1,1)]
{\label{f:fitted}\includegraphics[width=0.47\textwidth]{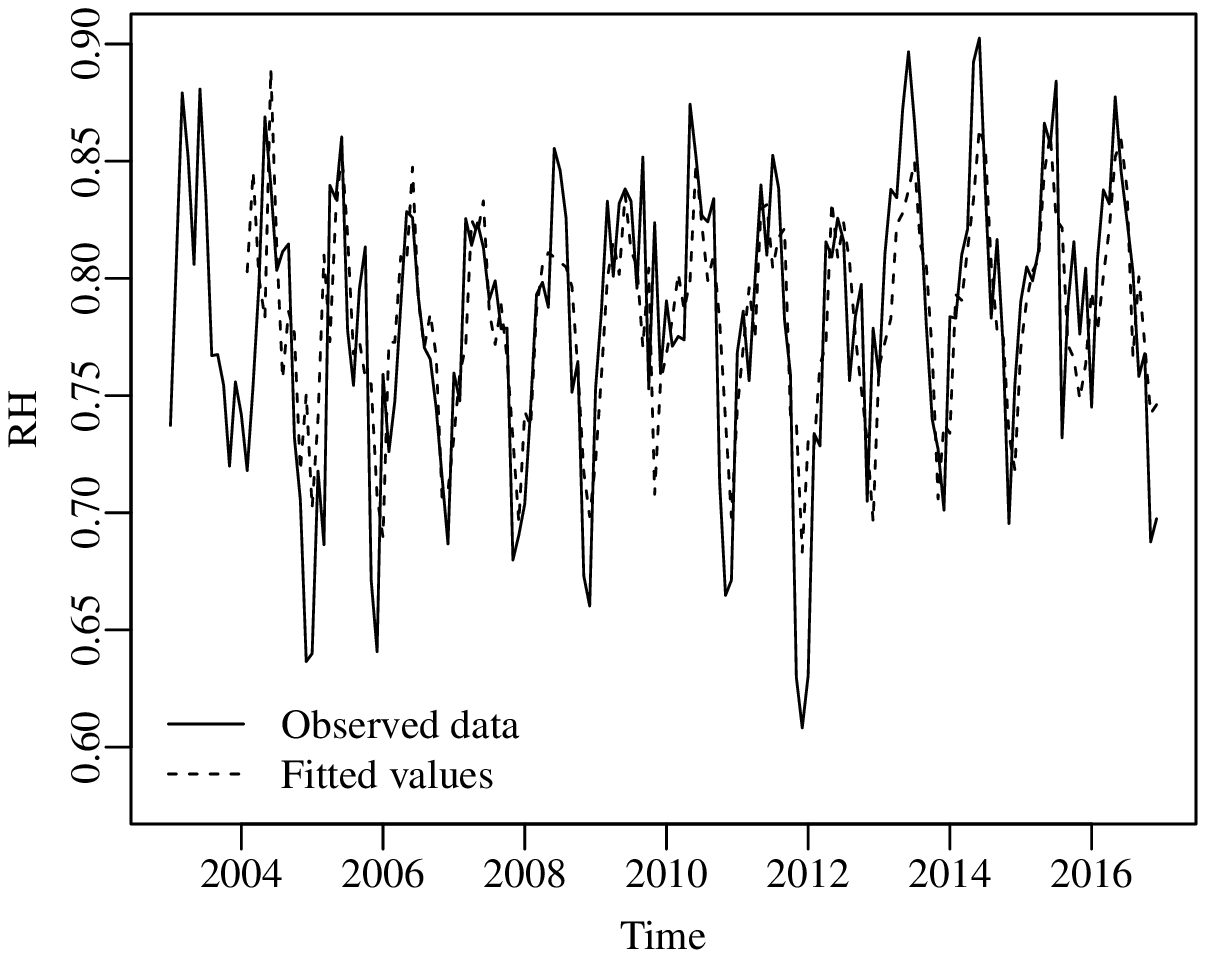}}
\subfigure[Observed and predicted values, January through October 2017]
{\label{f:comparison}\includegraphics[width=0.47\textwidth] {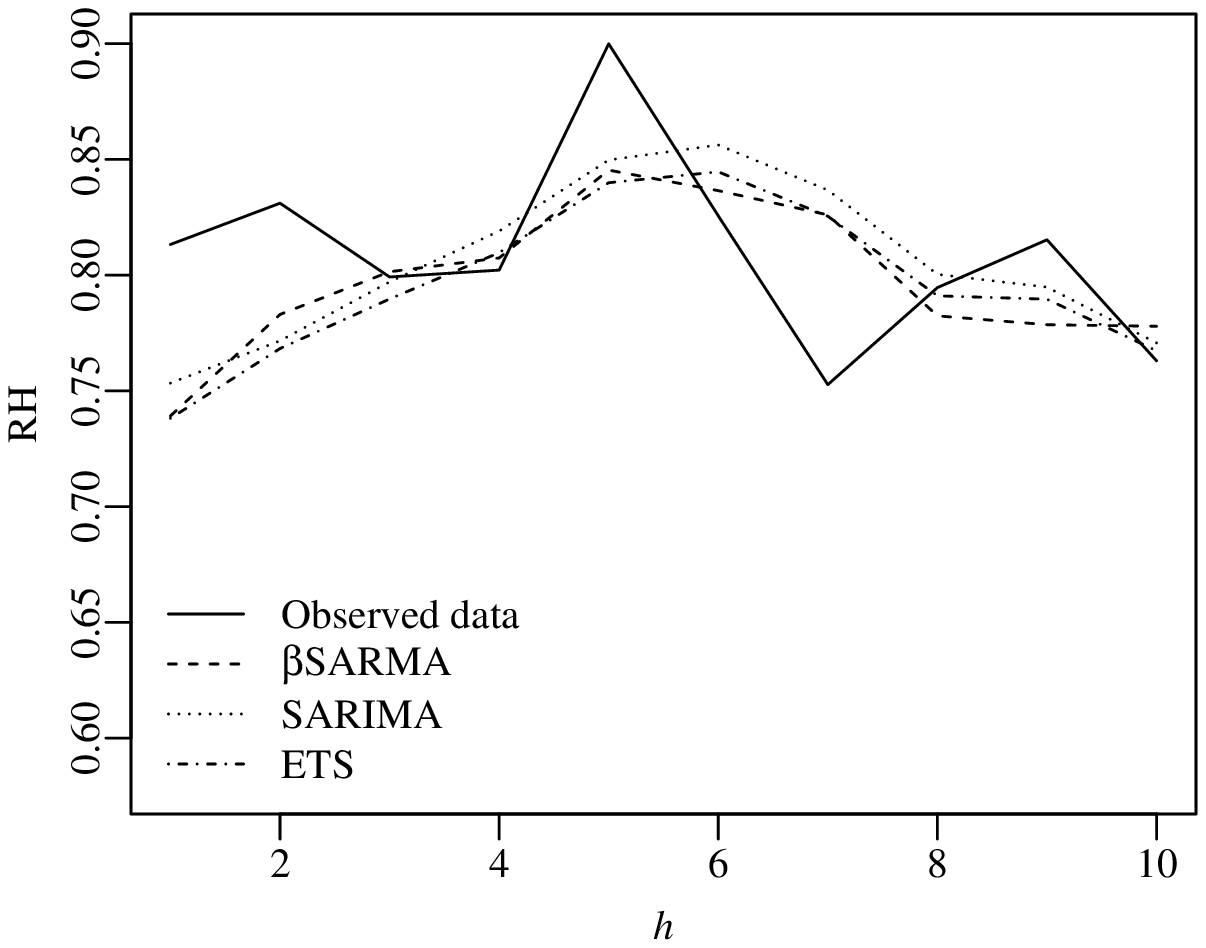}}
\caption{
Observed, fitted and predicted values of RH in Santa Maria, Brazil.
}\label{f:forecast}
\end{center}
\end{figure}

Figure~\ref{f:diag} contains six plots, namely: (a) observed versus fitted values, (b) index plot of the standardized residuals, (c) residual sample autocorrelation function, (d) residual sample partial autocorrelation function, (e) residuals QQ (quantile-quantile) plot, and (f) residual density estimate obtained using a Gaussian kernel, which is plotted alongside the standard normal density. All plots and tests indicate that the fitted model can be safely used for out-of-sample forecasting.
In Figure~\ref{f:fitted} we plot the data (solid line) together with in-sample predictions (dashed line).

We forecasted the next ten (out-of-sample) observations using the  $\beta$SARMA(1,0)$\times$(1,1) model, the SARIMA(1,0,0)$\times$(1,0,1) and also exponential smoothing state space models (ETS) \cite{Hyndman2008}. (Recall that such observations were removed from the data at the outset.) The SARIMA and ETS forecasts were produced using the \texttt{forecast} \textsc{R} package \cite{forecast}.
The three sets of out-of-sample forecasts are presented in Figure~\ref{f:comparison}.
Table~\ref{t:forecast} presents the mean square error (MSE) and the mean absolute percentage error (MAPE) computed for each methodology. It is noteworthy that the $\beta$SARMA forecasts are the most accurate according to both criteria.

\begin{table}%
\caption{MSE and MAPE for predicted values from different models} \label{t:forecast}
\begin{center}
\begin{tabular}{lcc}
\hline
Model & MSE	& MAPE \\
\hline
$\beta$SARMA	& $0.00180$	& $0.04094$ \\
SARIMA	& $0.00197$	& $0.04172$ \\
ETS	    & $0.00184$	& $0.04158$ \\
\hline
\end{tabular}
\end{center}%
\end{table}

\FloatBarrier

\section{Conclusions}\label{S:conclusions}

Oftentimes practitioners need to model and predict the future behavior of times series that assume values in the standard unit interval. The interest may lie, for example, in modeling the behavior of a rate (e.g., unemployment rate) or of a proportion over time. Such time series dynamics may be impacted by seasonal fluctuations. In this paper, we introduced the class of seasonal $\beta$ARMA models, $\beta$SARMA. It generalizes the class of $\beta$ARMA processes and can be used to model and predict time series that assume values in the standard unit interval and are subject to seasonal fluctuations. We showed that parameter estimation can be carried out by conditional maximum likelihood. We derived closed-form expressions for the score vector and for the conditional information matrix. Interval estimation, hypothesis testing inference and model selection were also covered. We presented three different residuals that can be used to assess goodness-of-fit and two white noise noise tests that can be applied to the residuals computed from the fitted model. We also provided Monte Carlo evidence on the finite sample accuracy of point estimation and of two white noise tests. An empirical application was presented and discussed.

\section*{Acknowledgments}

We thank an anonymous referee for comments and suggestions. We also gratefully acknowledge partial financial support from CNPq/Brazil.

\bibliographystyle{ieeetr}
\bibliography{betareg}

\end{document}